\documentclass[12pt]{article}
\usepackage[dvips]{epsfig}
\usepackage{amssymb}

\newcommand{\beq}{\begin{equation}}
\newcommand{\eeq}{\end{equation}}
\newcommand{\bq}{\begin{quotation}}
\newcommand{\eq}{\end{quotation}}
\newcommand{\bc}{\begin{center}}
\newcommand{\ec}{\end{center}}
\newcommand{\BFACE}[1] {\mbox{\boldmath $#1$} }

\begin{document}

\title{{\vspace{0cm}
\sc On the physical basis of cosmic time}}

\author{
{\sc S.E. Rugh\footnote{Symposion, 
`The Socrates Spirit', Section for Philosophy and the Foundations of Physics,
Helleb\ae kgade 27, Copenhagen N, Denmark
({\em e-mail: rugh@symposion.dk)}} 
\addtocounter{footnote}{5}
and  H.
Zinkernagel\footnote{Department of Philosophy, Granada University, 18071
Granada, Spain ({\em e-mail: zink@ugr.es}).}} }
\date{}

\maketitle

\begin{abstract}
\noindent
In this manuscript we initiate a systematic examination of the physical
basis for the time concept in cosmology. We discuss and defend the idea
that the physical basis of the time concept is necessarily related to
physical processes which could conceivably take place among the material
constituents available in the universe. As a consequence we motivate the
idea that one cannot, in a well-defined manner, speak about time
`before' such physical processes were possible, and in particular, the
idea that one cannot speak about a time {\em {scale}} `before'
scale-setting physical processes were possible. It is common practice to
link the concept of cosmic time with a space-time metric set up to
describe the universe at large scales, and then define a cosmic time $t$
as what is measured by a comoving standard clock. We want to examine,
however, the physical basis for setting up a comoving reference frame
and, in particular, what could be meant by a standard clock. For this
purpose we introduce the concept of a `core' of a clock (which, for a
standard clock in cosmology, is a scale-setting physical process) and we
ask if such a core can---in principle---be found in the available
physics contemplated in the various `stages' of the early universe. We
find that a first problem arises above the quark-gluon phase transition
(which roughly occurs when the cosmological model is extrapolated back
to $\sim 10^{-5}$ seconds) where there might be no bound systems left,
and the concept of a physical length scale to a certain extent
disappears. A more serious problem appears above the electroweak phase
transition believed to occur at $\sim 10^{-11}$ seconds. At this point
the property of mass (almost) disappears and it becomes difficult to
identify a physical basis for concepts like length scale, energy scale
and temperature -- which are all intimately linked to the concept of
time in modern cosmology. This situation suggests that the concept of a
time scale in `very early' universe cosmology lacks a physical basis or,
at least, that the time scale will have to be based on speculative new
physics.
\end{abstract}

\newpage


\newpage

\section{Introduction}

Most cosmologists would agree that the physics describing the `material
content' of the universe becomes increasingly speculative the further we
go back in time. By contrast, it is widely assumed that the concept of
time (and space) itself -- by virtue of a cosmological space-time metric
-- can be safely extrapolated sixty orders of magnitude back from the
present to the Planck scales. Apart from some interesting hints in
Misner, Thorne, and Wheeler (1973) (see also Misner 1969), we have
found no discussions in cosmology which address the issue of whether
time, like the physical description of the material content, could
become more and more speculative as we go back in `time'. Studies
addressing the time concept at the Planck scale are of course
abundant, cf. the problem of time in quantum gravity and quantum
cosmology. But what we want to question here is whether the time concept
is well-defined as a physical concept in cosmology `before' (in the
backward extrapolation from the present) the Planck scale is reached.
The guiding question is thus: How far back in time can we go while
maintaining a well-defined time concept? 

It is standard to assume that a number of important events took place in
the first tiny fractions of a second `after' the big bang. For instance,
the universe is thought to have been in a quark-gluon phase between
$10^{-11}-10^{-5}$ seconds, whereas the fundamental material
constituents are massless (due to the electroweak (Higgs) transition) at
times earlier than $\sim 10^{-11}$ seconds. A phase of inflation is
envisaged (in some models) to have taken place around $10^{-34}$ seconds
after the big bang. A rough summary of the phases of the early universe
is given in the figure:

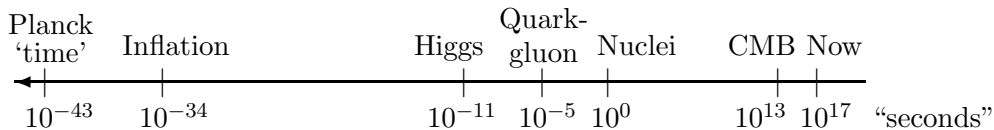
\begin{figure}[h]
\small
\label{timeline}
\setlength{\unitlength}{.87cm}                                                
\begin{picture}(3,3)(-1.5,-.3)

\thicklines

\put(12.8,1){\vector(-1,0){13}}

\put(.2,.88){{\bf $|$}}
\put(-.3,1.3){{\shortstack {Planck \\ `time'}}}
\put(-.05,.3){{$10^{-43}$}}

\put(2,.88){{\bf $|$}}
\put(1.45,1.4){{Inflation}}
\put(1.7,.3){{$10^{-34}$}}

\put(6.6,.88){{\bf $|$}}
\put(5.9,1.4){{Higgs}}
\put(6.1,.3){{$10^{-11}$}}

\put(7.8,.88){{\bf $|$}}
\put(7.2,1.3){{\shortstack{Quark- \\ gluon}}}
\put(7.5,.3){{$10^{-5}$}}

\put(8.8,.88){{\bf $|$}}
\put(8.7,1.4){{Nuclei}}
\put(8.6,.3){{$10^{0}$}}

\put(11.4,.88){{\bf $|$}}
\put(10.7,1.4){{CMB}}
\put(10.8,.3){{$10^{13}$}}

\put(12,.88){{\bf $|$}}
\put(11.9,1.4){{Now}}
\put(11.8,.3){{$10^{17}$}}

\put(12.9,.3){{``seconds"}}

\end{picture}
\end{figure}

\noindent While the various phases indicated in this figure will be
discussed in some detail in the present manuscript, a few comments and
clarifications should be made here: 

(i) The figure is to scale, that is, it captures e.g. that it is
(logarithmically) {\em shorter} from the present back to the Higgs
transition -- which more or less indicates the current limit of known
physics (as explored in Earth-based experiments) -- than from the Higgs
transition back to the Planck time located at $(\hbar G/c^5)^{1/2} \sim
10^{-43}$ seconds. This illustrates just how far extrapolations extend
in modern cosmology!\footnote{Prior to $\sim 10^{-2}$ seconds `after'
the big bang (the beginning of primordial nucleosynthesis) there is no
clear-cut {\em observational} handle on physics in the cosmological context,
see e.g. Kolb and Turner (1990, p. 74). The gap between this point and
the Planck time spans 41 orders of magnitude. (After the COBE and WMAP
experiments, however, it is widely believed that inflationary models may
have observational signatures in the cosmic microwave background
radiation (CMB)).} 

(ii) Whereas one usually speaks of time elapsed {\em since} the big
bang, the observational point of departure is the present -- hence the
direction of the arrow (we extrapolate backwards from now). For lack of
viable alternatives, however, we shall in the following use the standard
time indications from the big bang (we shall thus also speak about
`seconds after the BB').

(iii) The quotation marks around seconds are included since, as we shall
discuss, it is far from straightforward that one can `carry back' this
physical scale as far as one would like.
 
An objection to the study we propose might be that if time is well-defined 
within the Friedmann-Lema\^{\i}tre-Robertson-Walker (FLRW)
metric, standardly taken to describe the present universe (at large
scales), there seems to be no problem in extrapolating this time concept
back to $t=0$ or, at least, to the Planck time. 
However, this objection disregards that the
FLRW metric is a mathematical model containing a parameter $t$ which is
{\em interpreted} as time. Whereas, as a mathematical study, one may
consider arbitrary small values of $t$, our aim here is precisely to
investigate under what conditions -- and in which $t$-parameter range --
one is justified in making the interpretation
$$
t \; \; \leftrightarrow  \; \; \mbox{time.}
$$

In this paper we shall motivate and discuss the suggestion that a
physical condition for making the $t \leftrightarrow$ time
interpretation in cosmology is the (at least possible) existence of a
physical process which can function as what we call the
`core' of a clock. In particular, we suggest that in order to make the
$t \leftrightarrow$ time interpretation at a specific cosmological
`epoch', the physical process acting as the core of a clock should 1)
have a well-defined duration which is sufficiently
fine-grained to `time' the epoch in question; and 2) be a process which
could conceivably take place among the material constituents available
in the universe at this epoch. Consequently, we shall devote a large
part of the investigation to an examination of what such a core of a
clock could be in the context of early universe cosmology. Our analysis
suggests that the physical basis of time -- or, more precisely, the time
scale -- becomes rather uncertain already when the FLRW metric
is extrapolated back to $\sim 10^{-11}$ seconds. This could indicate
that the time scale concept becomes insufficiently founded (or at least
highly speculative) already $\sim 30$ orders of magnitude `before' the
Planck time is reached.
 
Our reasoning is based on the observation that we shall be (almost)
unable to find scale-setting physical processes (cores of clocks) in the
`desert' above the Higgs phase transition -- if the physics is based on
an extrapolation of what is considered well-known and established
physics in the form of the standard model of the electroweak and strong
forces. In order to provide a physical foundation for the time scale
above the Higgs transition we will have to base it on speculative new
physics, and the time scale linked to this new physics will be
speculative as well. Moreover, the in-principle existence of extended
physical objects which can function as rods (which appear to be a
prerequisite to set up the coordinate frame in cosmology,
see section 3) becomes gradually less clear: Above the quark-hadron
phase transition (at $t \sim 10^{-5}$ seconds) there are roughly no
bound systems left, and the notion of length and time scales becomes
even more ill-defined above the Higgs phase transition (at $t \sim
10^{-11}$ seconds) if those scales are to be constructed out of the
by-then available massless material constituents. 

The structure of the paper is as follows. In section 2 we discuss the
meaning of time, and suggest that the well-defined use of time in both
ordinary practical language and physics is necessarily related to the
notion of a physical process which can function as a clock or a core of
a clock. In section 3 we briefly investigate the time and clock concepts
as they are employed in cosmology and the underlying theories of
relativity.\footnote{The present manuscript focuses on the classical
space-time description of the universe and the possibility to identify
physical processes which can function as (cores of) clocks. Although
time is a classical parameter also in quantum physics, aspects of the
problem about time which are directly related to the quantum nature of
the physical constituents of the universe will be examined in a separate
investigation, see Rugh and Zinkernagel (2008).} In section 4 we examine
the possible physical underpinnings for (cores of) clocks in the early
universe. Results from this analysis are employed in section 5 where we
discuss how the identification of (cores of) clocks becomes
progressively more problematic as we go to smaller $t$-values in the
FLRW metric. A summary and some concluding remarks are offered in a
final section.

\section {The meaning of time}
\label{philosophy}

The concepts of time and space are so fundamentally interwoven in our
daily and scientific language that it is difficult to extract an
unambiguous meaning (or definition) of these concepts. In the
present manuscript we shall restrict our investigation to an examination of the
time concept in the realm of modern cosmology in which our ordinary
language (and its refinements in modern physics theory) is pressed to
the utmost. In this section we try to establish a few general points on
the meaning of time, which are relevant to cosmology.\footnote{In this
manuscript we are exclusively concerned with the meaning of time in
physics and what may be called ordinary practical language, e.g. we do
not address psychological, poetic or religious uses of the concept. The
body of literature dealing with the concept of time in general is rather
large. J.T. Fraser has estimated that the number of potentially relevant
references (books and articles from 1900 to 1980) for a systematic study
of time is around 65,000. (cf. Fraser and Soulsby, ``The literature of
time", pp. 142 - 144 in Whitrow (2003)).} 
 
It has been known at least since St. Augustine that it is difficult, if not
impossible, to give a {\em reductive} definition of time in terms of
other concepts that are themselves independent of time (see e.g. Gale
1968, pp. 3-4). For instance, operationalism, `time is what a clock
measures', is ruled out as a reductive definition, since one cannot
specify what a clock (or a measurement) means without invoking temporal
notions. Also it seems difficult to provide a meaning of time via a
referential theory of meaning, assuming e.g. that 'time is (or means)
whatever it corresponds to in reality', since one cannot point to or
specify what this referent might be (or at least, as we shall discuss
below, it is difficult to do this unless the referent somehow involves
clocks). The difficulties in providing a meaning (or definition) of time
point to the common intuition that time is a fundamental concept, i.e.
it is not possible to define the concept of time reductively.

Even though we cannot provide a reductive definition of time, we are
nevertheless able to use, and understand, the concept of time in a
non-ambiguous manner in a variety of situations in both ordinary
practical language and in physics. This suggests that the meaning of
time might be extracted from some common characteristic feature (or
features) of the use of the time concept in these
situations.\footnote{Of course, `meaning' is a contentious philosophical
notion. We are here assuming, inspired also by ideas in Zinkernagel
(1962), that a relevant alternative to operational and referential
meaning theories for time is a version of Wittgenstein's idea that the
meaning of a concept is somehow given by its use.} In both ordinary
language and in physics we use expressions such as ``this event takes a
certain amount of time", ``this event occurred such and such amount of
time ago", or ``this event is earlier/later than that event". In these
expressions, and in a multitude of other expressions involving time, the
meaning is usually non-ambiguous since the `time of the event' or the
`amount of time' can be identified by referring to some physical process
which can function as a clock (or, a `core' of a clock, see below).
Indeed, as far as we can see, these (and similar) expressions involving
time could not be given an unambiguous meaning {\em except} by referring
to some physical process. For instance, how should ``3 years" be
understood in a sentence like ``3 years passed, yet no physical process
took place (or, counterfactually, could have taken place) at all"? If
reference to physical (clock) processes is not presupposed then how is
such a statement to be distinguished from a similar one involving 4, 5
or any other number of years? As indicated above, however, (physical
processes which can function as or in) clocks cannot define time
reductively since any analysis of what a clock is (and does) will at
some point involve temporal notions. We therefore venture to formulate
the following time-clock relation -- which is an attempt to extract a
thesis concerning the correct use of the time concept from ordinary
practical language and physics:

\begin{quote} {\bf The time-clock relation:} There is a logical (or
conceptually necessary) relation between `time' and `a physical process
which can function as a clock (or a core of a clock)' in the sense that
we cannot -- in a well-defined way -- use either of these concepts
without referring to (or presupposing) the other. \end{quote}

We denote this relation `time-clock' in part to avoid the more
cumbersome `time-physical process which can function as
a clock (or a core of a clock)', and in part to follow the standard practice in
cosmology textbooks where time is associated with so-called `standard
clocks' (see section 2.3) -- usually without any specification of what
such standard clocks are. In accordance with this labelling, we shall in
the following often use `clock' as shorthand for `physical process which
can function as a (core of a) clock'. 

What is a core of a clock? According to a common conception, an ordinary
clock consists of a system which undergoes a physical process, as well
as some kind of counter (e.g. a clock dial) which registers increments
of time.\footnote{The second component of a(n) (ordinary) clock, that of
an irreversible registration with a counter, connects the discussion of
the concept of time with the existence of `an arrow of time' (an issue
we shall not pursue here).} Note that a change in the counter (e.g. a
displacement of the hands in a grandfather clock) is itself a physical
process which can be said to function as a clock -- as opposed to the
swinging pendulum inside the grandfather clock which functions as the
core of the clock.\footnote{Either one of these processes
(hand-displacement and pendulum-swing) may serve to make certain
temporal statements well-defined, hence the formulation `...clock or
core of a clock' in the time-clock relation.} Since no real
functioning clocks (with counters or dials) are available in the early
universe we shall in the following focus on physical processes which can
function as cores of clocks. As we shall discuss below, the requirements
that a physical process must satisfy in order to function as a core of a
clock will depend on which characteristic of time -- temporal order or
temporal scale -- is in question.\footnote{It is standard to distinguish
between order and metrical aspects of time, cf. e.g. Newton-Smith
(1980). With a temporal order one may speak of some event being before
or after another event (however, as we do not discuss the arrow of time,
our notion of time order refers not to the direction of time but merely
to the idea that two events are temporally separated -- that is, either
before or after one another). The notion of a temporal metric allows for
addressing quantitatively how much time separates two events. In this
manuscript we shall be particularly interested in the time {\em{scale}}
(a metrical aspect of time) which assigns specific values to the
duration of time intervals in some chosen unit.} 

The time-clock relation does not imply that physical processes are more
fundamental than time but is rather a thesis stating that time and
physical processes which can function as (cores of) clocks cannot be
defined independently of one another. Although formulated as a relation
between concepts, the time-clock relation reaches beyond the conceptual
level.\footnote{The notion of logical relations between concepts (which
in formal logic are relations between concepts like `not', `and',
`both-and' etc.) is inspired by P. Zinkernagel (1962 and 2001).} For
instance, the relation is also meant to capture the idea that the use of
the time concept, e.g. as in the above example with ``3 years", refers
to (actually or counterfactually) existing physical processes. In this
sense the use of the time concept presupposes both the possible (actual
or counterfactual) {\em{existence}} and the {\em{concept}} of physical
processes (see also the discussion in Zinkernagel 2008).

\subsection{Implicit definition of time via laws of nature}

By possible physical processes we shall understand processes which are
allowed by the laws of nature. This connects the time-clock relation
with the idea of an implicit definition of time via laws. To appreciate
this connection, we first note that although laws of nature (which are
typically evolutionary laws with respect to a time parameter $t$) can
refine, or make precise, our pre-theoretical notion of time, they cannot
provide a {\em {reductive}} definition of time in terms of other
concepts appearing in the laws, since at least some of these concepts
will themselves depend on -- and have to be specified as a function of
-- time; see below for an example. More importantly, the idea of
implicitly defining time via laws is in conformity with the time-clock
relation since, in our assessment, any such definition must make use of
actual or counterfactual clocks -- or, more precisely, of some possible
physical process (or class of processes) which can
function as (clocks or) cores of clocks.

Consider for instance Poincar\' e's analysis of the link between the
time concept and the mathematical structure of the natural laws (1905,
p.215): "So that the definition implicitly adopted by the astronomers
may be summed up thus: time should be so defined that the equations of
mechanics may be as simple as possible".\footnote{The idea of implicitly
defining time via simplicity considerations of the dynamical equations
in question is also discussed in Brown (2006, p. 92-93).} This idea of
an implicit definition of time addresses not what time as such is
(again, it is not a reductive definition) but rather {\em {which}} time
parameter is the most adequate (or, in Poincar\' e's terminology, most
convenient) for the physical situation in question. Poincar\' e
discussed what later became known as ephemeris time: Observations of the
positions of the bodies in the solar system (in terms of the time
standard provided by the earth's rotation) lead to discrepancies with
the positions of these bodies as predicted by Newtonian mechanics -- and
ephemeris time is the time standard which minimizes the discrepancies,
and thus makes Newton's laws come out right (there will, however, be
residual effects e.g. due to general relativistic corrections and
limited observational accuracy; for more details see e.g. Barbour (1999)
p.106, and Audoin and Guinot (2001) p.46 ff). A simpler example is
mentioned by Misner, Thorne and Wheeler (1973, p. 26): In the case of
moving free particles, the time parameter should be chosen, in
accordance with Newton's first law, so as to make the tracks of the
particles through a local region of spacetime look straight (that is,
the particles should traverse equal distance in equal times) -- any
other choice of time will make the motion look complicated (the tracks
will be curved). 

The important point for both ephemeris time and free particle motion is
that in either case the implicit definition of time is made by appealing
not only to the laws but also to some physical clock system (or class of
physical clock systems): the solar system (an actual clock) in the case
of ephemeris time, and free particles (an ideal, and hence
counterfactual, clock) in Misner, Thorne and Wheeler's
case.\footnote{Note that although there is a difference between
realistic (more or less accurate) clocks and perfect (or ideal) clocks,
this difference is not important for the time-clock relation: perfect
clocks are never realized but we may still refer to such clocks as
idealizations over real clocks, and we can, at least in principle,
estimate the inaccuracy of any clock using the laws of nature if we know
about the imprecision involved (this is e.g. how we determine that even
the best atomic clocks are slightly imperfect). Thus, as far as the
time-clock relation goes, any clock can be used to make time a
well-defined concept. For instance, it is a perfectly meaningful
statement to say ``according to my old wrist-watch, the football match
lasted only 88 minutes". Of course, in physics and astronomy one is
often after good (close to perfect) clocks, since these are the most
useful ones.}

The same conclusion can be reached if we attempt to make the implicit
definition of time via laws `explicit' by expressing time as a function
of the other physical quantities appearing in the laws. Consider, for
instance, how the time parameter $t$ could be extracted from the
`evolution' of the other physical quantities which appear in Newton's
second law, $\BFACE{F} = m (d^2\BFACE{r}/dt^2)$. Given two sets of values
of $\BFACE{F}$ and $\BFACE{r}$ (assuming $m$ is constant) one might
attempt to determine the time (duration) it takes for the system to get
from $(\BFACE{F}_1, \BFACE{r}_1)$ to $(\BFACE{F}_2, \BFACE{r}_2)$ by
inverting and integrating (twice) Newton's law and solve for $t$.
However, in order to perform the integration and get a definite value,
one would have to specify (a further initial condition, e.g. the initial
velocity $\BFACE{v}_1$ and) the force field $\BFACE{F}$ as a function of
$\BFACE{r}$ and $t$. Otherwise, one cannot determine the trajectory
connecting $(\BFACE{F}_1, \BFACE{r}_1)$ and $(\BFACE{F}_2, \BFACE{r}_2)$
and thus the time it takes for the system to get from one state to the
other. This not only reaffirms that one needs a notion of time from the
outset (so that time cannot be reductively defined by the law), but the
resulting picture is also in conformity with the time-clock relation:
Specifying the force field acting on the mass in question is equivalent
to specifying a physical system (or class of systems) --- such as a free
particle, $F=0$, a simple harmonic oscillator, $F=kr$, or a simple
pendulum in a constant gravitational field, $F=mg\sin\phi$ --- and the
process bringing the system from $(\BFACE{F}_1, \BFACE{r}_1)$ to
$(\BFACE{F}_2, \BFACE{r}_2)$ is a physical process which can function as
the core of a clock. Also in this case, then, the `implicit' definition
of time is made by appealing to possible physical (clock) systems.

With this clarification of the close connection between the time-clock
relation and natural laws in mind, we can now specify what we take to be
implied by the statement `time has a well-defined use'. According to the
time-clock relation, it involves a reference to some physical process (a
process in conformity with physical laws). This can also be expressed by
saying that the well-defined use of time requires that the process
referred to has a physical basis. In the following, we shall formulate
this idea as the assertion that a well-defined use of time requires that
time has a physical basis. More specifically, for the time concept to
have a physical basis we shall require one of two things (see also below
and section 4) -- depending on whether it is the time scale or the time
order (or both) which is in question: The time scale has a physical
basis if some actual or possible physical process has a duration shorter
or equal to the time interval in question.\footnote{This means that we
would consider a time interval of, say, $10^{-100}$ seconds a purely
mathematical abstraction (with no physical basis) insofar as no physical
process (in conformity with known laws) which could exemplify such an
interval is thinkable.} The time order, i.e. that an event A is before
(or after) an event B, has a physical basis if some actual or possible
physical process can take place between A and B. Of course, in ordinary
practical language and in physics we, in general, need a physical basis
(and a well-defined use) of both the time scale and the time order.

\subsection{The time-clock relation and relationism}

Since clocks involve change, the time-clock relation is in conformity
with a version of relationism according to which time is dependent on
change. This kind of relationism is in the tradition of Aristotle who
argued in his {\em Physics} (350 B.C.) that time is the measure (or
number) of motion, and that time and motion define each other. Also
Leibniz argued that time must be understood in relation to matter and
motion: ``Space and matter differ, as time and motion. However, these
things, though different, are inseparable" (Leibniz 1716, cf. Alexander
(1956, p. 78)). Nevertheless, the relationism implied by the time-clock
relation diverges from both Aristotle's and Leibniz's versions. First,
as already noted, the time-clock relation allows counterfactual clocks
(or motion) and is thus at odds with Aristotle's relationism insofar as
his version refers exclusively to actual (measured) motion. Second, we
do not argue, as did Leibniz, for a reductive relationism -- i.e. that
time can be reduced to temporal relations between events. To assume that
time is just temporal (before, after, and simultaneous) relations
between events leaves open what exactly a {\em temporal} relation is (or
what is temporal about the relations). We doubt that there is any way to
explain what `temporal' (and `before', `after', and `simultaneous')
means without invoking the notion of time itself.\footnote{For instance
we doubt that temporal relations can, as suggested e.g. by Leibniz,
Reichenbach, and Gr{\"{u}}nbaum, be reduced to causal relations in a
non-circular manner.} If this doubt is justified such a reductive or
eliminative analysis of time cannot work.

The non-reductive aspect of the relationism defended here sets us apart
from some modern relationists who otherwise also argue for the necessity
of clocks in order to give meaning to (at least the metrical aspects of)
time. Thus, for instance, Gr{\"{u}}nbaum (1977) advocates a relationism
in the context of general relativity which holds that ``[i]n a {\em
non}empty, metrically structured space-time, metric standards [like rods
and clocks] external to this space-time do play an ontologically
constitutive role in its very metricality or possession of any metric
structure at all, and the gravitational field of {\em empty} space-times
are devoid of {\em geometrical} physical significance..." (p. 341,
emphasis in original). Another example, inspired by Leibniz and with a
view to quantum gravity, is Smolin (1999, p. 238) who argues that
``...at least at the present time, the only useful concept of time that
exists in general relativity, in the cosmological context, is ... the
time as measured by a physical clock that is part of the universe and is
dynamically coupled to the rest of it". A related example is Barbour
(1999). If one abstracts from Barbour's (quantum gravity inspired) goal
of eliminating time altogether, his relationism has similarities to what
we are defending, cf. e.g. ``...time is told by matter -- something has
to move if we are to speak of time" (1999, p. 100).\footnote{Barbour's
advocacy of relationism in classical mechanics involves the derivation
of the time metric from considering the dynamical system in question in
a way similar to the way ephemeris time is defined (Barbour 1999, pp.
104 ff.). And, as Barbour notes (1989, p.181) ``...ephemeris time is
only abstract in the sense that it is not realized by any one particular
motion but is concrete in the sense that it must be determined
empirically from actually observed motions". This might well be an
indication that Barbour would agree with the point made above, viz. that
the definition of ephemeris time, and other implicit definitions of time,
appeals both to (Newton's) laws and to a specific physical (clock)
system.}

We are aware that the time-clock relation, and the relationism
accompanying it, is likely to be rejected by substantivalists (and
presumably by some relationists as well).\footnote{In general,
substantivalists hold that (space-)time exists independently of material
things, see e.g. Hoefer (1996) for a classification and discussion of
various versions of substantivalism. In light of the theories of
relativity, with their notion of four dimensional space-time, the debate
between relationism and substantivalism is mostly about {\em both} space
and time. In this manuscript we shall mainly focus on time, and we do
not claim that our considerations about time and clocks hold analogously
for space and rods. For instance, while we would argue that physical
(and metrical) space does need material systems which could function as
rods to be interpreted as such, we shall not here question the space
concept {\em far away} from us (or in the `empty space' inside a
spatially extended material box) whereas we do try to question (see
below) the time concept {\em long time ago} (in the `very early'
universe).} However, we take it be a difficult challenge for
substantivalists and other critics of the time-clock relation to account
for the meaning, or well-defined use, of time independently of clocks.
For, in our view, such an alternative account of time ought to be
closely related to our usual notion of time in ordinary practical
language and physics (which does depend on clocks) -- otherwise why call
it time at all?

For instance, a manifold substantivalist (who holds that space-time is
exhaustively described by a four dimensional manifold, see e.g. Earman
and Norton (1987)) will presumably hold that the meaning of time can be
inferred from, or identified with, a mathematical structure (the
manifold). In our assessment, however, while time can be (and in physics
always is) represented mathematically, it would be a mistake to think
that there is nothing more to time than a mathematical structure. The
fact that our usual notion of time can be associated with a formal
(mathematical) structure originates in time being related to physical
(clock) systems, e.g. as when we count the number of
sunrises and represent the amount by natural numbers. As mentioned
above, it is this association to physical systems which is used to give
unambiguous meaning to statements involving time e.g. statements about
the temporal separation between events. In any case, as noted e.g. by
Hoefer (1996, p. 11) there is nothing inherent in a (mathematical)
manifold which distinguishes space and time.

The substantivalist could more plausibly hold that time is defined as a
part of the space-time `container' constituted by a manifold plus a
metric, in which physical objects and processes are (or may be) embedded
in accordance with physical laws.\footnote{This corresponds to what
Hoefer (1996) calls manifold plus metric substantivalism or metric field
substantivalism (the distinction between these substantivalisms is
important for responses to the famous hole argument, but makes no
difference for our discussion).} If this view is understood as implying
that time is implicitly defined by laws of nature it is not in conflict
with the time-clock relation since (i) as discussed above, implicit
definitions of time rely on possible physical processes; and (ii) the
time-clock relation refers to actual {\em {or}} counterfactual physical
processes in conformity with the laws of nature. It might seem that the
inclusion of the metric, subject to Einstein's field equations, in what
is called space-time could give a physical basis for both the time order
{\em and} the time scale independently of any matter content or material
processes (e.g. by referring to processes involving only gravitational
waves). However, the manifold and the metric (and the vacuum Einstein's
equations) by themselves cannot provide a time (or length)
scale since neither $c$ nor the combination of $c$ and $G$ sets a time
(or length) scale. This means that unless the material content,
represented by $T_{\mu\nu}$ on the right-hand side of Einstein's
equations, provides a physical scale (see section 4) there is no
physical basis for a time scale. Essentially the same point was made
already by Eddington ((1939, p. 76), see also Whitrow (1980, p. 281)):

\begin{quotation}
\noindent
...relativity theory has to go outside its own borders to obtain the
definition of length, without which it cannot begin. It is the
microscopic structure of matter which introduces a definite scale of
things.
\end{quotation}

\noindent
As we shall see, the (micro-) physical laws operative in the very early
universe (and thus the associated energy-momentum tensor $T_{\mu\nu}$)
may not be able to provide the required physical scales. If this is the
case then one can at most address the question of time {\em {order}} and
not time {\em {duration}} for this `epoch' (see also below).

\subsection{The role of clocks in cosmology}
\label{three}

The time-clock relation is in conformity with the use of time in
cosmology although cosmologists often formulate themselves in
operationalist terms -- that is, invoking observers measuring on factual
clocks.\footnote{Our view on the time concept represents a departure
from operationalism in several ways: Apart from the point that clocks
cannot define time reductively (since clocks and measurements depend on
the time concept), we allow reference to counterfactual clocks; we
attempt to construct the {\em cores} of clocks out of available physics,
but do not require that this core should be associated with a counter
mechanism that could transform it into a real functioning clock; and we
do not require the existence of observers and actual measurements.} For
instance, Peacock (1999, p. 67) writes concerning the
Friedmann-Lemaitre-Robertson-Walker (FLRW) model, which is the standard
cosmological model (and the one we discuss in the present manuscript):

\begin{quote}
We can define a global time coordinate $t$, which is the time measured
by clocks of these observers -- i.e. $t$ is the proper time measured by
an observer at rest with respect to the local matter distribution. 
\end{quote}

\noindent
While this reference to clocks (or `standard' clocks) carried by 
comoving observers is widely made in cosmology textbooks, there is
usually no discussion concerning the origin and nature of these clocks.
Part of the motivation for the present investigation is to provide a
discussion of this kind.

Whereas cosmologists often refer to clocks as sketched above, they also
define cosmic time `implicitly' by the specific cosmological model
employed to describe the universe (e.g. through the relation between
time and the scale factor, see section 4). However, in accordance with
the discussion above, a definition of cosmic time via the cosmological
model requires that the time concept of the model can be given a
physical basis. Moreover, a backward extrapolation of the time concept
of the model requires that the model itself can be given a physical
basis.\footnote{The question of what it means for a model to have a
physical basis connects to a vast literature in the philosophy of
science (for an introduction to this literature, see e.g. Frigg and
Hartman 2006). We take it that for a cosmological model to have a
physical basis involves, as a minimum, that the (physical) conditions to
set up the model are at least approximately satisfied in the domain in
which the model is applied. In fact, as we shall discuss further in
section 3, a more basic requirement for a cosmological model to have a
physical basis is that the concepts involved in the model (e.g. concepts
like "homogeneity", "particle trajectory" and, of course, "time") are
physically grounded in the domain in which the model is applied.
Otherwise, one could not, in a well-defined way, pose the question (let
alone determine the answer) of whether or not the physical conditions to
set up the model are satisfied.} 

The FLRW model has a built-in $t$ concept which, as a mathematical
study, can be extrapolated back arbitrarily close to zero. The question
is whether the mathematical parameter $t$ can be given the physical
interpretation of time for {\em all} (positive) values of $t$, or only
for some (more) restricted subset of the $t$-interval. The $t
\leftrightarrow \mbox{time}$ interpretation in the FLRW model can be
made at present since the physical conditions for setting up the FLRW
model are (approximately) satisfied at present (section 3) and since the
$t$ parameter in the model can be correlated to physical (clock)
processes. For instance, clocks on earth measure time intervals
approximately in accordance with $t$ intervals in the FLRW model since
earth clocks are approximately at rest in the comoving frame (see
section 3). Given the $t \leftrightarrow \mbox{time}$ interpretation of
the FLRW model at present, one might then imagine extrapolating the
physical interpretation of $t$ `all the way back in $t$', or at least
sixty orders of magnitude from the present back to the Planck scale. In
this way, the fact that the FLRW model has a physical basis in some $t$
parameter range is used to give meaning to time (via this model) in
other $t$ intervals. 

However, since the procedure in question involves the extrapolation of
the FLRW model and its time concept it requires in our view that
 the physical basis of time in the model and, more
generally, the physical conditions needed to set up the model, are not
invalidated along this extrapolation. As we shall see, a major problem
is that length and time scale setting physical processes permitted by
physical laws at present, may not be allowed in a much earlier cosmic
epoch. Note in this connection that the laws of physics are often
assumed to be invariant under translations in space and time (see e.g.
Ellis 2006). Hence one might think that if the time concept has a
physical basis in one (temporal) domain then it ought to have it in all
domains (modulo Planck scale considerations). In the cosmological
context however this time translation invariance of physical laws is
constrained since cosmological phase transitions may change the {\em
{effective}} laws of microphysics (in this way, one may speak of an
evolution in the laws of nature, see e.g. Schweber 
1997).\footnote{Note that since the concept of time may be seen to be
implicitly defined by the laws of physics (section 2.1) it is difficult
to say what we could mean by `time translation invariance of the laws of
physics'. With respect to which notion of time (living ``outside" the
realm of physical laws) should this time translation invariance of the
laws of physics be addressed? There is, however, a (widely accepted)
sense in which the effective laws change in cosmic time in the early
universe ($\rightarrow$ nuclear age $\rightarrow$ atomic physics age
$\rightarrow$ etc.) as the temperature decreases. This lack of time
translation invariance of the effective laws should be addressed with
respect to the $t$ parameter in the FLRW model. As we shall see, it is
precisely these effective laws which must be used to provide a physical
basis for (in particular, the scale of) the $t$ parameter in the FLRW
model, and thus these laws which allow the interpretation of $t$ as
(metrical) time.} In particular, the physical basis for the time scale
may break down in the very early universe as a result of phase
transitions, cf. section 5. 

The relevance of taking into consideration the physical basis of the
FLRW model and its time concept is exemplified by the widely accepted
expectation that the model cannot be extrapolated below Planck scales
and, accordingly, that the $t \leftrightarrow \mbox{time}$
interpretation cannot be made for $t$ values below $10^{-43}$
seconds.\footnote{The Planck scale limitation is inferred from the
expectation that the space-time metric (because of quantum effects)
fluctuates increasingly as the Planck scale is approached, and thus the
(classical) time concept is gradually rendered invalid. This
invalidation presumably affects both the order and the metrical aspect
of time insofar as no physical process can be clearly distinguished
`before' the Planck scale (cf. the problem of time in quantum gravity).}
This illustrates that a physical condition (namely that quantum effects
may be neglected) can imply a limitation for the $t \leftrightarrow$
time interpretation. But if it is accepted that there is at least {\em
one} physical condition which must be satisfied in order to trust the
backward extrapolation of the FLRW model and its time concept, it
appears reasonable to require that also {\em other} physical conditions
(which are necessary to set up the FLRW model) should be satisfied
during this extrapolation.\footnote{We note that -- except for the
space-time singularity itself -- there are no internal contradictions in
the {\em mathematics} of the FLRW model (or classical general
relativity) which suggests that this model should become invalid at some
point, e.g. at the Planck scale.} 

The idea of examining the $t \leftrightarrow \mbox{time}$ interpretation
in cosmology appears to be in the spirit of the discussion in Misner,
Thorne and Wheeler (1973) (henceforth MTW) concerning how present
physical scales might be extrapolated into the past. In connection with
the question of a singularity occurring at a finite past proper time
they write (p. 814):

\begin{quote}
``The cosmological singularity occurred ten thousand million years ago."
In this statement take time to mean the proper time along the world line
of the solar system, ephemeris time. Then the statement would have a
most direct physical significance if it meant that the Earth had
completed $10^{10}$ orbits about the sun since the beginning of the
universe. But proper time is not that closely tied to actual physical
phenomena. The statement merely implies that those $5 \times 10^9$ orbits
which the earth may have actually accomplished give a standard of time
which is to be {\em extrapolated in prescribed ways}, thus giving
theoretical meaning to the other $5 \times 10^9$ which are asserted to
have preceded the formation of the solar system. [our emphasis]
\end{quote}
 
\noindent But MTW does not explain what these ``prescribed ways" are. We
have suggested above that such backward extrapolation of physical scales
makes sense (has a physical basis) only insofar (i) it is still possible
to identify physical processes which may function as (the cores of)
clocks; and (ii) these clocks (based on the available physics) should
set a sufficiently fine-grained physical scale in order to specify the
time of the epoch. Perhaps MTW would have agreed, for they note (p. 814)
that even with better (and earlier) clocks there will always be a
residual problem concerning the physical notion of time:

\begin{quote} 
Each actual clock has its ``ticks" discounted by a suitable factor -
$3*10^{7}$ seconds per orbit from the Earth-sun system, $1.1*10^{-10}$
seconds per oscillation for the Cesium transition, etc. Since no single
clock (because of its finite size and strength) is conceivable all the
way back to the singularity, a statement about the proper time since the
singularity involves the concept of an infinite sequence of successively
smaller and sturdier clocks with their ticks then discounted and added.
[...] ...finiteness [of the age of the universe] would be judged by
counting the number of discrete ticks on {\em realizable clocks}, not by
accessing the weight of unrealizable mathematical abstractions.
\end{quote}

This quote seems to imply that the progressively more extreme physical
conditions, as we extrapolate the FLRW model backwards, demand a
succession of gradually more fine-grained clocks to give meaning to
(provide a physical basis of) the time of each of the 
epochs.\footnote{For instance, no stable Cesium atoms -- let alone real
functioning Cesium clocks -- can exist before the time of decoupling
of radiation and matter, about 380,000 years after the big bang.} In the
same spirit, our view is that a minimal requirement for
interpreting $t$ as time (with a scale) is that it must be possible to
find physical processes (the cores of clocks) with a specified duration
in the physics envisaged in the various epochs of cosmic
history.\footnote{As noted earlier, the time-clock relation is also satisfied
by counterfactual clocks, that is, even if no actual physical
(clock) process correlated with $t$ took place in an interval of cosmic
time, it is sufficient if some physical process {\em could possibly}
(consistent with the available physics) have taken place. For our
purposes in this paper this criterion is sufficient since, as we shall
argue, both actual and counterfactual clocks (which can set definite
time scales e.g. in the unit of seconds) may be ruled out by the physics
describing the very early universe. If one were to contemplate a
possible world with altogether different laws than in our universe,
allowing for different clocks -- ``counterlegal" from the point of view
of our universe -- then there might be no problem about very early times
(thanks to Jeremy Butterfield for bringing this possibility to our
attention). On the other hand, the relevance of this is not obvious as
it is not clear whether or how the $t$ (time) coordinates in the two
worlds would be related.} In section 4 and 5, we shall investigate in
more detail how far back in cosmic time this criterion can be satisfied.

In general, the duration of the physical process functioning as the core
of a clock will specify a standard of time but, as we shall see, there
might be epochs where it is only possible to speak of `scale free'
clocks, which can order events but not determine how much time there is
in between them. We note that if a point is reached in the backward
extrapolation of the FLRW model where no `scale clocks', but only
non-metrical `order clocks' could exist, one would lose all handle on
how close we are to the singularity. Since modern cosmology makes
extensive use of specific times, e.g. to locate the cosmological phase
transitions, we shall in the following focus on physical clocks which
can set a time scale. One may {\em speculate} on the existence of
(scale-setting, i.e. metrical) clocks based on ideas beyond presently
known physics. However, if such speculative clocks are invoked to give
meaning to some epoch in cosmology, it should be admitted that the time
of this `epoch' becomes as speculative as the contemplated physics of
these clocks.

\section {Time in Relativity and FLRW cosmology}
\label{TimeBB}

The conceptual apparatus of standard cosmology (and its interpretation)
originates within the framework of general relativity in which various
conceptions of time are at play. Rovelli (1995), for instance,
distinguishes between coordinate time (which can be rescaled
arbitrarily), proper time ($d\tau = ds/c = 1/c
\sqrt{g_{\mu\nu}dx^{\mu}dx^{\nu}}$), and clock time (which is what is
measured by a more or less realistic clock). By itself, coordinate
`time' has no physical content so one might argue that it should not
even be called time but rather just $t$. Coordinate `time' is however
identified with proper time in some models of GR (in which a `gauge'
fixing has constrained the choice of coordinate system, as in FLRW
cosmology -- see below). And proper time does deserve its name (time),
at least when it is associated with the time of a standard clock, cf.
also MTW (1973, p. 814):

\begin{quote} 
...proper time is the most physically significant, most
physical real time we know. It corresponds to the ticking of physical
clocks and measures the natural rhythms of actual events.
\end{quote}

\subsection{The necessity of rods and clocks in relativity}

The indispensability of (rods and) clocks for a physical interpretation
of relativity theory is not a novel idea. As is well known, Einstein
initiated the theory in a positivistic spirit which implied an
operational meaning criterion e.g. for the definition of time in terms
of what clocks measure. But even if Einstein subscribed less to a
positivistic philosophy over the years (Ryckman 2001), he kept referring
to the role of measuring rods and clocks in order to give physical
meaning to the coordinates in special relativity. By contrast, it is
often emphasized that there is no physical meaning to the coordinates in
general relativity (e.g Rovelli 2001).\footnote{Einstein's own remarks
on time in GR are somewhat obscure, see e.g. Einstein (1920), chap. 28,
where he suggests defining time in terms of `Dali like' clocks attached
to deformable non-rigid reference bodies (Einstein calls them
``reference-molluscs"). While Einstein suggests that the laws of motion
for these clocks can be anything, he does demand ``...that the
`readings' which are observed simultaneously on adjacent clocks (in
space) differ from each other by an indefinitely small amount" (Einstein
1920, p. 99). But this seems possible only if the clocks work normally
(as in special relativity) and in this sense are standard clocks.}
However, as we argue throughout this manuscript, the possible existence
of (the core of) rods and clocks -- i.e. length and time scale setting
physical objects or processes from microphysics -- are in any case
indispensable in GR (and the FLRW model) at least in order to provide a
physical foundation for the scales of length and time.

The necessity of {\em material} rods and clocks (external to GR itself)
can be, and has been, questioned e.g. by referring to the
`geometrodynamical clock' developed by Marzke and Wheeler (see e.g. MTW
p. 397-399) which does not depend on the structure of matter (see also
Ryckman 2001 for references concerning the limitations of the concept of
a rigid rod in general relativity). Thus, MTW writes (p. 396): ``In
principle, one can build ideal rods and clocks from the geodesic world
lines of freely falling test particles and photons. [...] In other
words, spacetime has its own rods and clocks built into itself, even
when matter and nongravitational fields are absent!". The
geometrodynamical clock is built out of two `mirrors' moving in parallel
between which a pulse of light is bouncing back and forth. The mirrors
can be any system absorbing and re-emitting light. In the cosmological
context, however, there are at least two problems in using such clocks
as a physical underpinning for the FLRW reference frame (for a general
discussion and critique of the Marzke-Wheeler construction, see Gr{\"
u}nbaum 1973, p. 730 ff.). First, mirror systems moving in parallel are
presumably difficult to envisage (even in principle) from the available
constituents in the early universe. Second, and more importantly, since
the electromagnetic theory of light is scale invariant such light clocks
cannot set a scale which can provide a physical basis for specific times
in cosmology. As we shall discuss later on, there might be `epochs' in
the early universe in which it is no longer possible to identify
physical systems (among the envisaged constituents) which may function
as (the core of) rods and clocks.\footnote{In
particular, we shall see in section 5 that electromagnetic theory
including sources (e.g. electrons) -- as well as the other constituents
of the standard model of particle physics -- is scale invariant above
the electroweak phase transition. This implies, for instance, that not
even the test particles (the mirrors) of the Marzke-Wheeler
geometrodynamical clock are able to set a length scale or a
time scale above (`before') this phase transition.}

\subsection{Time in big bang cosmology}

After the above brief remarks about time in relativity, we now turn to
cosmology. Time depends on the chosen reference frame
already within the limited set of coordinate transformations allowed in
special relativity. Thus, it is necessary to specify a privileged
coordinate frame in cosmology if a conception of cosmological time
(defined throughout the universe) is to be found. The large scale
space-time structure of the cosmological standard model is based on the
FLRW line element. Apart from a discussion on the exact choice of the
coordinates, the form of this quantity can be derived from two
assumptions, namely

\begin{enumerate}

\item Weyl's postulate. The world lines of galaxies (or `fundamental
particles') form a bundle of non-intersecting geodesics orthogonal to a
series of spacelike hypersurfaces. This series of hypersurfaces allows
for a common cosmic time and the spacelike hypersurfaces are the
surfaces of simultaneity with respect to this cosmic time (see
below).\footnote{ We note that the while the Weyl principle and the
cosmological principle allow for the possibility to set up a {\em
global} cosmic time, the implementation of these principles can only be
motivated physically if we already have a physical foundation for the
concepts of `space' and `time' {\em locally} -- otherwise we cannot
apply concepts like "spacelike", "spatially homogeneous", "spatially
isotropic", which appear in the definitions of Weyl's principle and the
cosmological principle.}

\item The cosmological principle. This states that the universe, on
large scales, is spatially homogeneous and spatially
isotropic.\footnote{The requirement of `isotropy' is to be understood
relative to the privileged reference frame outlined below (see also
Weinberg (1972, p. 410) and MTW (1973, p. 714)). Some cosmologists (e.g.
Pietronero and Labini (2004)) doubt whether observations actually
justify the homogeneity assumption at the largest observed scales ($\sim
800$ Mpc). In the present manuscript, however, we shall confine our
discussion to the cosmological standard model, and leave out possible
alternatives, see e.g., Narlikar (2002) and L\'{o}pez-Corredoira (2003)
and references therein.}

\end{enumerate}

\noindent The idea of the Weyl postulate is to build up a comoving
reference frame in which the constituents of the universe are at rest
(on average) relative to the comoving coordinates, cf. Narlikar (2002,
p. 107 ff.) (for a historical account of Weyl's postulate, or principle,
see North (1990, p. 100 ff.)). The trajectories $x_i = \mbox{constant}$
of the constituents are freely falling geodesics, and the requirement
that the geodesics be orthogonal to the spacelike hypersurfaces
translates into the requirement $g_{0i} = 0$, which (globally) resolves
the space-time into space and time (a 3+1 split). We have $g_{00} = 1$
if we choose the time coordinate $t$ so that it corresponds to proper
time ($dt = ds/c$) along the lines of constant $x_i$ (see Robertson
1933), i.e. $t$ corresponds to clock time for a standard clock at rest
in the comoving coordinate system. The metric is thereby resolved into $
ds^2 = c^2 dt^2 - g_{ij} (x,t) dx^i dx^j $.\footnote{As noted e.g. by
Coles (2001, p. 313), this synchronous coordinate system,
$g_{0\mu}=(1,0,0,0)$, is ``...the most commonly used way of defining
time in cosmology. Other ways are, however, possible and indeed useful
in other circumstances."} The spatial part of this metric is then
simplified considerably by application of the cosmological principle
(see below).

The empirical adequacy of both the Weyl postulate and the cosmological
principle depends on the physical constituents of the
universe.\footnote{For instance, Robertson (1933) noted that the
reintroduction in cosmology of a significant simultaneity (in the
comoving coordinate system) implied by Weyl's postulate is permissible
since observations support the idea that galaxies (in average) are
moving away from each other with a mean motion which represents the
actual motion to within relatively small and unsystematic deviations.}
We shall later observe that as we go backwards in time it may become
increasingly more difficult to satisfy, or {\em even formulate}, these
principles as physical principles since the nature of the physical
constituents is changing from galaxies, to relativistic gas particles,
and to entirely massless particles moving with velocity
$c$.\footnote{Narlikar (2002) comments (p. 131) that for a gas of
relativistic particles where one has random movement ``...the Weyl
postulate is not satisfied for a typical particle, but it may still be
applied to the center of mass of a typical spherical volume". Above the
EW phase transition, all constituents move with velocity $c$ (in any
reference frame) and, thus, there will be no constituents which will be
comoving (at rest) relative to the FLRW reference frame. If one would
like to construct mathematical points (comoving with the reference
frame) like the above mentioned center of mass (or, in special
relativity, center of energy) out of the massless, ultrarelativistic gas
particles, this procedure presumably requires that a length scale be
available in order to e.g. specify how far the particles are apart
(which is needed as input in the mathematical expression for the
``center of energy").} Correspondingly, as we shall see, the effective
physical laws describing the constituents of the universe change as the
temperature rises and eventually become scale invariant above the
electroweak phase transition. This makes it difficult to even formulate
e.g. the cosmological principle (let alone decide whether it is
satisfied): If an epoch is reached in which the fundamental constituents
become massless, so that a local scale invariance (conformal invariance)
of the microphysics is obtained (see section 5.4), it would seem that
the concept of homogeneity lacks a physical basis. For homogeneity rests
on the idea of (constant) energy density in different spatial points and
(constant) energy density is not an invariant concept under local scale
transformations.\footnote{Apart from the necessity of physically based
scales in order to set up the Weyl and cosmological principle, we may
also recall (while insisting that we are not defending operationalism!)
that it, of course, requires actual rods and clocks to establish these
principles empirically -- for instance it requires clocks to study the
redshifts (cf. section 4.1) of galaxies moving away from us and it
requires rods to probe the homogeneity of the matter distribution.} On
top of this, the physical basis of the Weyl postulate (e.g.
``non-intersecting world lines of fundamental particles") appears
questionable if some period in cosmic history is reached where the
`fundamental particles' are described by wave-functions $\psi (x,t)$
referring to (entangled) quantum constituents. What is a `world line' or
a `particle trajectory' then?\footnote{The problem of setting up a
cosmic time when quantum theory is taken into account is addressed in
Rugh and Zinkernagel (2008). One may also ask whether individual quantum
constituents, like a proton, are sufficient to function as rods needed
to set up the spatial part of the reference frame with well-defined
x-,y- and z- directions (even when quantum constituents introduce a
physical scale, such as the proton radius, it may be questioned whether
these constituents can point in a well-defined direction).}

After these critical remarks on the (limited) physical foundation of the
principles which underline the FLRW model, let us briefly
review how the standard big bang solution is obtained. The most general
line element which satisfies Weyl's postulate and the cosmological
principle is the FLRW line element (see e.g. Narlikar 2002, p. 111 ff.):

\beq
\label{lineelement}
ds^2=c^2dt^2 - R^2(t)\left\{ \frac{dr^2}{1-kr^2} + r^2 (d\theta ^2 +
\sin ^2 \theta d\phi ^2) \right\}
\eeq
where $R(t)$ is the scale factor, $k=0,-1, +1$ correspond, respectively,
to a flat, open and closed geometry, and $r$, $\theta$, and $\phi$ are
comoving (spherical) coordinates which remain fixed for
any object moving with the expansion of the universe. The
spatial part of the metric describes a curved space which is expanding
with cosmological proper time $t$ as described by the cosmic scale
factor $R(t)$. 

The symmetry constraints encoded in the FLRW metric require that the
energy-momentum tensor of the universe necessarily takes the form of a
perfect fluid. The fundamental equations which determine the dynamics of
the FLRW metric are the Einstein equations, an energy-conservation
equation and the equation of state (for the perfect fluid). Applying the
Einstein field equations to the FLRW metric results in two equations (the
Friedmann equations) which relate the energy and pressure densities to
the Hubble parameter ($H \equiv \dot{R}/R$, where the dot denotes a
derivative with respect to cosmic time $t$), see e.g. MTW p. 729. With
perfect fluid matter, pressure and energy densities are proportional, $p
= p(\rho) = w \rho$ (the equation of state), where e.g. $w=0$ for dust
(pressureless matter) and $w=1/3$ for a radiation gas. The evolution of
the scale factor $R = R(t)$ can be found by solving the Friedmann
equations. A particularly simple solution is obtained in the flat,
$k=0$, case assuming $w$ to be constant (see e.g. Coles and Lucchin
1995, p. 34): 

\beq
\label{rt-relation}
R(t) = R_0 \left( \frac{t}{t_0} \right)^{2/(3(1+w))}
\eeq

\noindent where $R_0$ and $t_0$ are, respectively, the present value of
the scale factor and the present age of the universe. This (and other)
solution(s) to the Friedmann equations illustrates (within classical
GR) the idea of a big bang universe expanding from a singularity $R = 0$
at $t = 0$.

As already indicated, the standard physical interpretation of the $t$
coordinate in the FLRW model is to identify it with the proper time of a
{\em standard clock} at rest in the comoving system, i.e. $(r, \theta,
\phi)= \mbox{constant}$. In our view, a standard clock (or a core of a
standard clock) in cosmology is any actual or possible scale-setting
physical process which takes place in the cosmic epoch in question. The
following two sections will address what these clock processes could be.

\section{Possible types of clocks in cosmology} 
\label{types}

Below we will sketch some possible early universe clocks (giving
standards of time). First we discuss the possibility of founding the
time concept on the general features of the FLRW model ($R$, $\rho$,
$T$). Then we look at some major candidates among the microscopic
constituents of the early universe which might, more independently of
the FLRW model, provide a physical basis for cosmic time.

In general, when we try to associate the parameter $t$ $(= t(\cal C)) $
with some physical process or a physical concept ${\cal C}$, we cannot
require that the concept ${\cal C}$ does not depend on $t$. Indeed,
insofar as time is a fundamental concept (cf. section 2), there are
(virtuous) circles of this kind in any attempt to specify the time
concept. However, we shall require that ${\cal C}$ is not just a
mathematical concept, but a concept which has a foundation in physical
phenomena (in which case we shall say that $\cal{C}$ has a physical
basis). In particular, in order for ${\cal C}$ to provide a
physical basis for time in a given cosmological epoch, these physical
phenomena should refer to scales based on physics which are available at
the epoch in question. Otherwise, if the required scales are more fine-grained 
than those available, the physical foundation of ${\cal C}$ will
be insufficient. This may be remedied by introducing speculative physics
with the required fine-grained scales, but in this case ${\cal C}$, and
the time concept $t$ $(= t(\cal C)) $ in the given `epoch', become as
speculative as the physics employed. 

Before looking at possible early universe clocks, we shall briefly
illustrate the idea of founding the time concept upon physics involving
well-known phenomena and  scales. An elementary clock
system involves the propagation of a light signal as well as a physical
length scale. Such a clock was also discussed by Einstein (1949,
p. 55):
\begin{quote}
The presupposition of the existence (in principle) of (ideal, viz.,
perfect) measuring rods and clocks is not independent of each other;
since a light signal, which is reflected back and forth between the ends
of a rigid rod, constitutes an ideal clock,... 
\end{quote}

One way to provide a physical basis for the length scale involved in an
Einstein clock is to refer to a spatially extended object of the same
size within well-known physics. By contrast, if the length scale
involved in the Einstein clock is smaller than physically based length
scales from known physics, this clock -- and the time concept based on
it -- will be speculative.\footnote{As mentioned 
above, if shorter time
intervals (shorter length scales) are needed to provide a physical basis
for the time of a cosmological epoch, one may introduce speculative
elements into physics. As we shall see in section 5, such speculative
physics is needed not just because the physical length scales within
known physics may become `too coarse' to found the time concept in
cosmology. Even worse, known physics suggests that {\em any} physical
length scale -- fine-grained or coarse -- may disappear at some point in
the very early universe.} Compare this with Feynman's (Feynman et al
1963, p. 5-3) discussion of short times in particle physics where he
comments on the by-then recently discovered strange resonances whose
lifetime is inferred indirectly from experiments to be $\sim 10^{-24}$
seconds (see also subsection \ref{Decay} below). To illustrate this time
scale, Feynman appeals to what seems to be an intuitive physical picture
since it (roughly) corresponds to ``...the time it would take light
(which moves at the fastest known speed) to cross the nucleus of
hydrogen (the smallest known object)". Feynman continues

\begin{quotation} \noindent What about still smaller times? Does ``time"
exist on a still smaller scale? Does it make any sense to speak of
smaller times if we cannot measure - or perhaps even think sensibly
about - something which happens in shorter time? Perhaps not.
\end{quotation}

\noindent Whereas the lifetime of strange resonances -- or the $Z^0$
particle (section \ref{Decay}) -- is rooted in physical phenomena (even
if extracted from an elaborate network of theory dependent
interpretations) we shall briefly look at Feynman's intuitive and
illustrative picture of a light signal crossing a proton, and question
whether this is a physical process which can function as a core of a
clock (i.e. a physical process with a well-defined duration which can
provide a physical basis for the corresponding time interval). The
spatial extension of a proton is a physically motivated length scale,
but its quoted value coincides with the proton's Compton wavelength (see
also section 5.2). If the proton is to function as a ``measuring rod"
and the light pulse is to propagate from one end of the proton to the
other then the light pulse needs to have wavelength components which are
smaller than the extension of the proton. But the light pulse will then
have sufficient energy to create proton-antiproton pairs and we will
therefore have pair production of ``measuring rods" (so does the light
pulse reach the other end?). Moreover, elementary processes in quantum
electrodynamics do not have well-defined space-time locations so a
precise duration of a photon travelling a given distance cannot be
specified (see e.g. Bohr and Rosenfeld (1933) and Stueckelberg (1951)).
Thus, a closer inspection into the physical `light-crossing-proton'
process reveals intricate quantum field theory phenomena which questions
whether this process is capable of providing a physical basis for the
time scale of $10^{-24}$ seconds.

It should be emphasized that even if $10^{-24}$ seconds could be taken
to correspond to a physical process (e.g. a light signal crossing a
proton), this subdivision of the second cannot be achieved in the very
early universe (`before' there are protons), and therefore this physical
process cannot provide a physical basis for $10^{-24}$ seconds `after'
the big bang. Indeed, one should distinguish between how far the second
can be subdivided at present (in terms of physical structure available
now) and how far back in $t$ the FLRW model can be extrapolated while
maintaining a physical basis for time (and temperature).\footnote{In
fact, the electroweak processes (which at present may provide a physical
basis for time scales of $10^{-24} - 10^{-25}$ seconds) take place at a
cosmic time (according to the standard FLRW model) $\gtrsim 10^{- 11}$
seconds after the big bang.}

\subsection{The cosmic scale factor $R$ as a clock} 

An important large scale feature of the cosmological standard model is
that the scale factor $R(t)$ is one-to-one related to the parameter $t$.
Thus if we could devise a clear physical basis for the scale factor $R$,
we could invert the mathematical formula $R = R(t)$ to give the
parameter $t = t(R)$ as a function of $R$, and in this way base a
concept of time on the scale factor of the cosmological model (see also
MTW 1973, p. 730). For the flat $k=0$ case, it is particularly simple to
integrate the Einstein equations with perfect fluid matter $p = w \rho$,
and we get (cf. equation \ref{rt-relation})

\beq
\label{skalafactor}
t = t_0 \left (\frac{R}{R_0} \right )^{3(1+w)/2}  
\eeq

\noindent where, for instance, the exponent is equal to $2$ ($w=1/3$)
for a radiation dominated universe.\footnote{The
precise functional relationship between $t$ and $R(t)$ depends on which
particles exist at the relevant temperature, and of their interaction
properties, see e.g. Narlikar (2002, p. 171) or Kolb and Turner (1990,
p. 65, Fig. 3.5).} For the scale factor to serve as a clock, however,
one needs to have some bound system or a fixed physical length scale
which does not expand (or which expands differently than the universe).
Otherwise, we are simply replacing one mathematical quantity ($t$), with
another ($R(t)$), without providing any physical content to these
quantities. Indeed, if everything (all constituents) within the universe
expands at exactly the same rate as the overall scale factor, then
`expansion' is a physically empty concept.\footnote{Eddington, for
example, emphasized the importance of the expansion of the universe to
be defined relative to some bound systems by turning things upside-down:
`The theory of the ``expanding universe" might also be called the theory
of the ``shrinking atom" (Eddington 1933, quoted from Whitrow 1980, p.
293). See also Cooperstock et al. (1998).} We can therefore
ask:\footnote{In this and the following section, we shall attempt to
clarify the discussion by emphasizing key observations in some of the
subsections.}

\begin{quotation}
\noindent
{\bf Observation $\# 1$:} 
If there are no bound systems -- and not least if there are no other
physically founded fixed length scales (cf. section 5.2-5.4) --
in a contemplated earlier epoch of  
the universe, is it then meaningful to
say that the universe expands? Relative to which length scale is the
expansion of the universe to be meaningfully addressed?
\end{quotation}

\noindent {\em Implementing a scale via boundary conditions.} \\
\noindent It is customary to put in a time scale and a relative length
scale for the scale factor as a starting (boundary) condition of the
cosmological model. For example we could employ as an input in the FLRW
model that the present age of the universe is $t \sim 14$ billion years
($t \sim 4 \times 10^{17}$ seconds) and set the present relative scale
factor ($R/R_0$) to unity (see e.g. Narlikar 2002, p. 136, or Kolb and
Turner 1990, p. 73). By relying on present physical input scales, it
thus seems that the scale factor in equation (\ref{skalafactor}) can
serve as a clock as far back in cosmic time as desired. This
possibility, however, will depend on whether the (relative) scale factor
has a physical basis during the backward extrapolation (cf. section 2
and below). 

The physical basis of the relative scale factor can for instance be
given in terms of red shifts ($1+z = R_0/R$). In turn the physical
meaning of (redshifted) light depends on the physical meaning of
frequency and wavelength, and, as discussed in section 4.3, this
requires the existence of relevant physical length scales (as
observation 1 indicates).\footnote{Similar remarks apply to the option
of using the density $\rho$ (rather than the cosmic scale factor $R$) as
a clock. In a radiation dominated universe (cf. Weinberg 1972, p. 538)
we have $t \sim (G \rho)^{-1/2}$. As in the case of the scale factor,
however, we need to have physical scales (for volume and energy) in
order for the energy density concept $\rho$ to function as a clock in a
given cosmological epoch. Without a fixed volume concept (that is a
scale for length, e.g. a non-expanding bound system), there is no
physical basis of an energy density (which changes with time).
Furthermore, the fixed scales need to be based on known physics if the
density clock is to be non-speculative.}

\subsection{Temperature as a clock} 

The relation between the FLRW $t$ parameter and the occurrence of the
various (more or less sharply defined) epochs, in which different layers
of physics (atomic, nuclear,...) come into play, is identified via a
time-temperature relation. In order to establish this relationship
between time and temperature one first needs to relate the energy
density to temperature. According to the black-body radiation formula
(Stefan-Boltzmann law) the energy density ($\rho$) of radiation, which
is the dominant energy contribution in the early universe, is connected
to the temperature $T$ as 

\begin{equation} \label{rhoT}
\rho = a T^4 \equiv \frac{8 \pi^5 k^4}{15 h^3 c^3} \; T^4
\end{equation} 

\noindent which gives an energy density $\rho \approx 4.4 \times
10^{-31} kg/m^3$ of the present $T \sim 2.7 $ K microwave background
(comprised of a nearly uniform radiation of photons), see e.g. Weinberg
(1972, p. 509 and 528).\footnote{To obtain the total (dominant) energy
density in the early universe one has to sum over all the
(spin/polarization states of the) species of relativistic particles
(photons, neutrinos, ...), see e.g. Narlikar (2002, p. 170-171). Thus
equation (\ref{rhoT}) will be multiplied by a factor $g_*$ equal to the
number of effective relativistic (those species with mass $m_i c^2 \ll k
T$) degrees of freedom, for example $g_* = 2$ (2 polarizations) for a
photon. The number $g_*$ increases considerably with temperature. At $T
\gtrsim 300$ GeV $\approx 10^{15}$ K all the species in the standard
model of particle physics (8 gluons, $W^{\pm}$, $Z^0$, 3 generations of
quarks and leptons, and the Higgs) are effective relativistic degrees of
freedom, and $g_* > 100$ (see also Kolb and Turner 1990, p. 64-65).}
Moreover, throughout most of the early history of the universe the
temperature $T$ is simply inversely proportional to the scale factor: $T
\propto 1/R = 1/R(t)$.\footnote{Such a relationship $T \propto 1/R(t)$
can be understood intuitively if we assume that a typical wavelength of
the black body-radiation is stretched as the universe expands and simply
scales as $\lambda \propto R(t)$ (i.e. proportional to the scale
factor). However, the input assumptions really involve the application
of standard equilibrium thermodynamics (thermal equilibrium is assumed),
some simplifying approximations, and the application of principles such
as covariant energy conservation and the constancy of the entropy in a
volume $R(t)^3$. See e.g. Weinberg (1972), Sec. 15.6.} Thus, by using
the relation $t=t(R)$ (equation \ref{skalafactor}) -- obtained by
integrating the Einstein equations for the radiation dominated FLRW
universe -- one may arrive at a relationship between $t$ and $T$.

The more exact relation between time and temperature depends however
crucially on the temperature range considered, since different particles
and different particle properties come into play at different
temperatures. Analytic solutions can often not be obtained and one has
to resort to a numerical calculation to `integrate' through eras
in which complicated physical processes take place, for example the era
corresponding to temperatures around $T \sim 6 \times 10^9$ K in which
the process of annihilation of $e^{-} e^{+}$ pairs takes place. (The
electron mass $m_e$ corresponds to a temperature of $T \sim m_e c^2/ k
\sim 6 \times 10^{9}$ K). This is a process in which the ratio $m_e/ k
T$ of the electron mass to the temperature plays a decisive role and a
complicated formula governs how to arrive at the relationship between
time $t$ and the temperature $T$, 
\begin{equation} 
t = - \int \left (\frac{8}{3} \pi G a T^4 {\cal E} (\frac{m_e}{k T})
\right )^{-1/2} \; \left (\frac{d T}{T} + \frac{ d {\cal S} (m_e/k T)
}{3 {\cal S} (m_e/k T) } \right )
\end{equation} 
where ${\cal E} (x)$ and ${\cal S} (x)$ are two quite involved
mathematical functions of the argument $x = m_e/ k T$ (Weinberg 1972,
p. 537-540).

When the temperature of the universe is well above the $e^{-} e^{+} $
annihilation temperature $T \sim 6 \times 10^{9}$ K (but below the
quark-hadron phase transition, as well as the $\mu ^- \mu ^+$
annihilation temperature of $\sim 1.2 \times 10^{12}$ K), the
relationship between time $t$ and temperature $T$ simplifies
considerably (Weinberg 1972, p. 538):\footnote{Similar relations with
different prefactors also hold in other intervals of temperature where
the effective number of relativistic degrees of freedom remains
constant, cf. Kolb and Turner (1990, p. 64-65); Narlikar (2002, p.
170-171).}

\begin{equation} \label{timetemp}
t = \left (\frac{c^2}{48 \pi G a T^4 }\right )^{1/2} \sim 1 \mbox{
second} \times \left (\frac{T}{10^{10} \; K} \right )^{-2}
\end{equation}
According to this equation, starting at $T = 10^{12}$ K, it took roughly
$\sim 10^{-2} $ seconds for the temperature to relax to $10^{11}$ K, and
roughly $\sim 1$ second for the temperature to drop to $10^{10}$ K. To
establish time-temperature relations similar to equation
(\ref{timetemp}), as the temperature $T$ increases and the FLRW model is
extrapolated further backwards, requires knowledge of the particle
species in existence and of their properties. \\

\noindent
{\em The physical basis for temperature } \\
In order to employ temperature as a clock 
one needs to provide an account of the 
physical basis for the concept $T$ which appears in relations such as
(\ref{timetemp}). The concept of temperature is roughly a
measure of the average energy of the particles (the available energy per
degree of freedom). In a non-relativistic limit this energy is in turn
connected to the velocity of the particles, $ k T \propto \langle 1/2 m
v^2 \rangle $. Thus, in order for the temperature concept to have
physical meaning one needs a well-defined mass, time, and length scale.
For ultra relativistic particles (like massless photons) which move with
the velocity of light, $c$, the temperature concept is related to the
average energy density which demands a well-defined energy and volume
scale. We can thus state:

\begin{quotation}
\noindent
{\bf Observation $\# 2$}:
The temperature concept is intimately connected to length and energy (or
mass) scales. Therefore one cannot have a physical basis of the
temperature concept when such scales are not available. Moreover, if the
available scales are not of relevant size (not sufficiently fine-grained), 
the temperature concept will be insufficiently founded. If
scales of the relevant size are provided by speculative physics, the
temperature concept will be speculative as well.
\end{quotation}

We should also recall that temperature is a statistical concept which
has no meaning for individual particles but only for ensembles: There
must be a concept of (nearly) local thermodynamic equilibrium in order
to speak about a temperature, see also Kolb and Turner (1990, p. 70
ff.). The standard definition of temperature in thermodynamical
equilibrium reads 
\beq
1/T = (\partial S/ \partial E)_V
\eeq
where $S$ is the entropy and $E$ the free energy, and the volume $V$ is
kept fixed. It is again clear from this definition that a length scale
(volume scale) and an energy scale are presupposed (in sections 5.3 and
5.4, we will question this presupposition in the `desert epoch' above
the Higgs phase transition at $T \approx 300$ GeV $\sim 10^{15}$ K).

\subsection{The black body radiation clock}

In connection with the possibility of a temperature clock, we shall now
examine if one can use the frequency of the radiation (the massless
constituents) of the universe to provide a physical basis for the
concept of temperature or, directly, the concept of time. In
thermodynamical equilibrium at a temperature $T$ there is a distribution
of the photons among the various quantum states with definite values of
the oscillation frequencies $\omega$ as given by the Planck distribution
(black body radiation). The average oscillatory frequency grows
proportionally to the temperature $\hbar \omega = k T$ (where $k$ is the
Boltzmann constant translating a temperature into an energy scale).

Could we envisage taking a large ensemble of photons\footnote{Or any
particle species whose mass is much smaller than $k T$ so that it can be
considered effectively massless and treated on the same footing as
photons.} in the early universe and extract the average wavelength or
oscillation frequency of
the ensemble?\footnote{Since frequency depends on the chosen reference
frame, we might imagine some fictitious rest frame, say, in which the
the momenta $\hbar \BFACE{k}$ of all the photons sum up to zero.}
One could then imagine converting the average frequency $\nu$
into a time scale $t \sim 1/ \nu$, and in this way
establishing a physical time scale at the temperature $T$. This would be
a promising proposal for a `pendulum' (oscillation frequencies of
massless particles) which abounds in the universe and which could
possibly even function down to Planck scales, thus providing a physical
basis to the concept of time (or temperature) right down to Planck
scales.

However, just like in the above examples, this frequency clock relies on
physically based input scales (in order for $\nu$ to have a physical
basis). One may attempt to use the radiation field to establish a time
scale directly, without referring to an external length scale, by
counting the number of oscillations of a test charge which is located in
the electromagnetic radiation field. But how many oscillations should we
count in order to put a mark that we now have $1$ second (or $10^{-40}$
seconds)? If the oscillation frequency is not linked to some physical
process setting a scale it appears downright impossible to convert a
number of oscillations into a time standard (given in any fixed time
unit).\footnote{Similar remarks apply, of course, to the physical
significance of the wavelength $\lambda = c \times 1/\nu$ of the
electromagnetic wave. Such a wavelength is lacking any physical basis if
there is no length scale setting process to which it can be related.}
One would need, for example, photons with a well-defined frequency
produced, say, in an annihilation process of particles with a given rest
mass or simply refer to a specified temperature scale, since the
oscillation frequency (by the black-body radiation formula) is linked to
temperature. Such a procedure would take us back into the discussion of
the above subsection 4.2.\footnote{The situation can be illustrated by
an analogy. The frequency for small oscillations of a pendulum of length
$\ell$ in a homogeneous gravitational field is given by $ \omega =
2 \pi / T = \sqrt{g/\ell}$. It is, however, impossible to relate a
counted number of oscillations of this pendulum to, say, the scale of a second 
unless we know the value of $g/\ell$. In general, we need (extra) input
physics in order to relate a counted number of oscillations to any fixed time
unit.}

The fact that electromagnetic light is not able, by itself, to set a
length scale can also be expressed by the observation (section 5.3) that
the massless standard model (which includes the electromagnetic
interaction) is scale invariant.

\subsection{Decay of unstable particles as clocks}
\label{Decay}

We proceed in this subsection to give a short discussion of whether
the decay processes of some material constituents in the universe could
provide clocks.

If we have $N_0$ identical particles and we suppose for a moment that we
can trace and follow the behavior of each of these, then we can imagine
counting the number of remaining $N (= N(t))$ particles after some time
$t$. In quantum theory the decay of an individual particle is described
by probabilistic laws so we need a statistical averaging procedure to
obtain a well-defined decay time (life time) $\tau$. From the number $N$
of remaining particles one may then estimate the time $t$ passed as 

\begin{equation} 
\label{decay}
t = \tau \; \times \; \log(N_0 / N) 
\end{equation} 

\noindent In this sense, the number $N$ of remaining particles functions
as a (statistical) clock. Given that we are interested in processes
which might function as clocks in the very early universe, it would be
most relevant to find decay processes which proceed very fast. For
example, we could imagine considering e.g. the decay of the muons $
\mu^{-} \rightarrow e^{-} \bar{\nu}_e \nu_{\mu}$, or of the $Z^{0}$
particles, $Z^0 \rightarrow f \bar{f}$, which can decay into any pair of
fermions.\footnote{The fastest process, within experimental particle
physics, is the estimated decay time, $\sim 10^{-25}$ seconds, for the
decay of a massive $Z^{0}$ particle with mass $M_Z \sim 91.1876 \pm
0.0021 \; GeV/c^2$ (the $Z^{0}$ is one of the three massive mediators of
the weak interactions). Note that such a tiny interval of time has not
been directly observed, but is inferred indirectly from resonance widths
and the Heisenberg uncertainty relation: The quoted lifetime can be
estimated (as $\sim \hbar/\Gamma$) from the measured resonance width
$\Gamma = 2.4952 \pm 0.0023$ GeV (see the ``Gauge \& Higgs Boson Summary
Table", p. 31 in S. Eidelman et al. (2004)).}

Two problems should be kept in mind (and further examined) if one would
like to use such (actual or counterfactual) decay processes as a
physical basis for time or the time scale in the early universe: (1)
Decay processes are quantum mechanical and therefore statistical in
nature. There is thus a minute probability that the number of particles
will remain constant for a `$t$ interval' (or even grow if particles are
created, say, by the opposite process $e^{-} \bar{\nu}_e \nu_{\mu}
\rightarrow \mu^{-} $). One has to check whether this means that a time
concept based on the decay of particles could stop or even go backwards
in these two cases (cf. Rugh and Zinkernagel 2008). (2) As mentioned
above, the statistical nature of the decay clock also means that we need
an averaging procedure to obtain a well-defined time scale. But in which
reference frame is this averaging procedure (in principle) to take
place? The expected decay time for an individual particle equals the
lifetime $\tau$ {\em only in the rest frame} of the decaying particle.
With respect to the comoving reference frame (in which cosmic time is
defined and in need of a physical basis, cf. section 3.2) the expected
decay time for each particle will be multiplied by the Lorentz factor
$\gamma$ which depend on the velocity of the decaying particle. This
velocity in turn depends on the temperature $T$. Thus, the decay time
relative to the comoving reference frame grows (linearly) with
temperature $T$, and an independent physical basis for temperature is
required in order for the decay clock to fix a well-defined scale. In
any case, all known decay clocks appear to be doomed to fail above the
electroweak phase transition since at this point the $\mu ^-$, $Z^{0}$,
and the entire spectrum of massive particles, become massless (see
section 5.3). 

Even if there are no decays, there may still be {\em interactions} among
the massless (charged) constituents of the ultrarelativistic plasma
which is envisaged to exist above the electroweak (Higgs) phase
transition. Since the physics is invariant under scale transformations
(section 5.4) such processes will, however, not by themselves be able to
set a fixed scale for time and length (at most, it may be possible for
such processes to provide order clocks). One may, for instance, imagine
introducing a `statistical clock' based on the interaction rate $\Gamma$
of the particles in the relativistic plasma.\footnote{This clock was
suggested to us by Tomislav Prokopec.} If we are following a particle,
and each time it interacts we count it as a `tick', we will have a
statistical clock in which the length of the ticks will fluctuate around
some mean value ($\tau = 1/\Gamma$), and the mean free path ($\ell = c
\tau$) sets a length scale (a definition which, for example, does not
rely on the notion of a bound system).\footnote{Note that this clock
needs a clear and unambiguous definition of the concept of `interaction
of two (massless) particles'. Otherwise how do we decide when two
particles interact? Is it when two point particles coincide in the same
space-time point (this involves renormalization problems, and a
prescription of how to identify `the same space-time point' which do not
involve rods)? Is it when the particle becomes `asymptotically free'
after a collision, cf. the S-matrix formalism (then when do we count)?,
Or is it when the particle is $x$ meters away from the collision partner
(then we need rods, or a length scale again)? The problem is probably
not eased if one takes into account the quantum nature of the particles
(entanglement).} The interaction rate can be estimated in the early
universe to be $\Gamma \sim \alpha kT/\hbar$, where $\alpha$ is the
relevant coupling constant, e.g. 1/137 for electromagnetism, and $T$ is
the temperature. In fact since $\tau = 1/\Gamma \sim 1/T$ the
(interaction rate) clock ticks slower and slower as the universe
expands. We note that the time and length ($\ell = c \tau$, the mean
free path) set by this clock scales exactly as the cosmological scale
factor $R \propto 1/T$. This entails (1) that the clock is not able to
set a fixed length scale relative to which the expansion of the universe
(expansion of the scale factor $R$) can be meaningfully addressed; and
(2) that the clock requires a physical foundation for temperature $T$ if
the tick-rate is to be employed as a clock which can provide a physical
basis for cosmic time scales given in some fixed unit of time (e.g. a
second).\footnote{More generally, in order to know that the basic tick
rate is changing we need some physical quantity that breaks the
(conformal) scaling symmetry. }

\subsection{Atomic (and nuclear) clocks}

In the realm of atomic and nuclear physics there are many physical
systems which have characteristic and well-defined oscillations and
periodicity. A photon produced by a transition between two well-defined
quantum states with energies $E_1$ and $E_2$ yields a well-defined
frequency establishing a characteristic time scale $\omega^{-1} =
\hbar/(E_1 - E_2)$.\footnote{One should not think of these frequencies
as being totally sharply defined, as there is always a line width (i.e.
an uncertainty $\bigtriangleup \omega$).} If we consider the simplest
bound system in atomic physics, the hydrogen atom, it has a spatial
extension of $\sim 10^{-10}$ m (the Bohr radius) whereas it emits light
with wavelengths of many hundred nanometers. Thus, the time scale of the
transitions (between quantum states) in a bound system like the hydrogen
atom is much longer (i.e. more `coarse') than the time scale it would
take light to cross the spatial extension of the system. The latter
would correspond to the ``Einstein light clock", mentioned in the
introductory remarks of section 4, for the case of light traversing the
spatially extended system of a hydrogen atom.

The spectral lines of hydrogen, the simplest bound atomic system, were
explained by Niels Bohr in 1913, and since then a wide range of quantum
mechanical transitions have been identified within atomic and nuclear
physics. The unit of a second is presently defined by
referring to a physical system within atomic physics: One second is the
duration of 9 192 631 770 periods of the radiation corresponding to the
transition between two well-defined hyperfine levels of the ground state
of the cesium-133 atom.\footnote{In any atomic clock, such as the
cesium-133 clock, one needs two components. (1) A frequency calibration
device: The frequency of a classical electromagnetic radiation signal is
calibrated to maximize the probability of a certain hyperfine transition
in a cesium atom. The procedure involves a large ensemble of cesium
atoms and the precision is limited by the statistics. (2) A counter:
There has to be a mechanism which can count 9 192 631 770 periods of the
radiation and thus (irreversibly) mark `1 second' when such a number has
been counted. See e.g. C. Hackman et al. (1995) or C. Audoin et al.
(2001).} 

Even if there were atomic systems before the `atomic age' of the
universe (from $\sim 10^{13}$ seconds after the big bang -- the time of
the emission of the cosmic microwave background radiation -- and
onwards), it is clear that there are no atoms before nucleosynthesis
(and also clear that there could not, counterfactually, have been any --
given the physical circumstances). Indeed, any atomic
clock will melt at a certain temperature and the use of atomic systems
as a physical basis for the time concept beyond this melting temperature
is thus excluded.

\section{The parameter $t$ in cosmic `history'}
\label{History}

In this section we shall attempt more explicitly to examine the $t$
concept, and the possibility to identify the `core' of (comoving)
clocks, within the various contemplated epochs of cosmic history. A key
parameter which determines the characteristic physics in each epoch is
the temperature $T$ since it is the transition temperatures of various
phase transitions (in the realm of nuclear and particle physics) which
give rise to the quoted times $t$ of each cosmic epoch.\footnote{A quick
summary of the contemplated epochs (eras) in cosmology may be found e.g.
in Coles (2001, p.348). Cf. also e.g. Weinberg (1972, Sec. 15.6) and
Narlikar (2002, Chapt. 6.).} We emphasize again that when we refer to a
time `after the big bang' (after the BB) in the following subsections,
we are (for lack of viable alternatives) following the standard
convention by which $t$ is `counted forward' from a fictive mathematical
point (the singularity in the FLRW model).

\subsection{Nucleosynthesis $\sim 10^{-2} - 10^2$ seconds after the big bang}

Since their inception in the late 1940's, primordial nucleosynthesis
models have yielded predictions of the relative abundances (measured
relative to the abundance of hydrogen) of various light nuclei. These
estimates are based on a detailed understanding of the possible nuclear
reactions among light nuclei which can take place at different
temperatures.\footnote{Agreement with astronomical observations is
obtained by fitting the baryon to photon ratio $\eta$ within the range
$\eta = n_B / n_{\gamma} \sim (4 \pm 2) \times 10^{-10}$, see, e.g.
Schramm and Turner (1998).}

According to the standard cosmological model, the synthesis of the light
nuclei is established over a time span of some minutes. The span of
cosmic time from $t \sim 10^{-2}$ to $10^2$ seconds after the BB is
estimated (via the FLRW solution) from the interval of cosmic
temperatures (from $\sim 10^{11} K \sim 10$ MeV to $\sim 10^9 K \sim
0.1$ MeV) at which the various primordial nucleosynthesis processes take
place. The cosmic time span available to the various nuclear reaction
processes is crucial for this scenario to produce the observed
abundances of light nuclei.\footnote{For example, the time span of the
processes relative to the neutron lifetime is crucial for the production
of helium. The free neutron decays in the weak decay $n \rightarrow p^+
+ e^- + \bar{\nu}$ and has a lifetime $\tau_n \sim 15$ minutes. If the
neutron lifetime were much shorter, the neutrons would have decayed
(within the available cosmic time of $\sim 10^2$ seconds) before
deuterium could form (and the helium production would never get
started).} Thus, in the era of cosmological nucleosynthesis there is
some (even observational) handle on the conception of cosmic time, and
there are time-scale setting nuclear processes on which we may base
(cores of) clocks in this era. The nucleosynthesis era
therefore provides an example in which cosmic time scales in early
universe cosmology are correlated with important time scales in rather
well understood micro-physical processes (within the realm of nuclear
physics).

\subsection{The quark-hadron phase transition: $\sim 10^{-5}$
seconds} 

Before the time of nucleosynthesis, the universe is believed to have been in
the lepton era in which the dominant contribution to the energy density
comes from electrons, electron-neutrinos, and other leptons (positrons,
muons, etc.). Even earlier, the universe is envisioned to have entered the
hadron era in which the material content is mainly comprised of hadrons,
such as pions, neutrons, and protons. 

A main focus of our discussion is the -- in principle -- availability of
(sufficiently fine-grained) length and time scale setting physical
objects or processes (cores of rods and clocks) at the given cosmic
moment in question. For this discussion a most important event takes
place at a transition temperature of $T \sim 10^{12}$ K ($\sim 10^{-5}$
seconds after the big bang). At this point, one imagines (as we
extrapolate backwards toward higher temperatures) that the hadrons
dissolve, their `boundaries' melt away, and the quarks break free. So
the universe is comprised of structureless particles -- quarks, leptons,
gluons, photons -- in thermal equilibrium. Above this so-called
quark-hadron transition one imagines the creation of `the quark-gluon
plasma', a hot and dense mixture of quarks and gluons, see e.g. Peacock
(1999, p. 305).\footnote{Before the establishment of the quark model in
the 1970's, cosmologists were reluctant to even speculate about times
earlier than $10^{-5}$ seconds, cf. e.g. Kolb and Turner (1990, p. xiv):
``...it was believed that the fundamental particles were leptons and
hadrons, and that at a time of about $10^{-5}$ sec and a temperature of
a few hundred MeV the strongly interacting particles should have been so
dense that average particle separations would have been less than
typical particle sizes, making an extrapolation to earlier times
nonsensical".}

Above the quark-hadron phase transition there are no bound systems left.
Below we shall discuss various physical properties of the microscopic
structure of matter which, apart from bound systems, may set a physical
length scale. As we shall see, it is not clear if such length scales as
the Compton wavelength (associated with a massive particle) and the de
Broglie wavelength (associated with the momentum of the particle)
constitute physical length scales which, e.g. together with a light
signal, can function as the core of a clock (a physical process with a
well-defined duration). Thus it may not be unreasonable to consider the
following possibility:

\begin{quotation} \noindent {\bf Observation $\#$ 3:} If a bound system
is a minimal requirement in order to have a physical
length scale (which can function in the core of a clock),
then it becomes difficult to find a physical basis for the
time scale above the quark-gluon transition at $t \sim 10^{-5}$
seconds.\end{quotation}

The idea of liberation of quarks in the quark-hadron phase transition is
widely accepted but there is no consensus as concerns the details of
this transition and the physical structure of the quark-gluon plasma;
see, for example, reviews by McLerran (2003) and M\"{u}ller and
Srivastava (2004). In particular, there is no consensus as concerns the
physical structure of the quark-gluon plasma phase. The transition
temperature for the quark-hadron phase transition is set by the
$\Lambda_{QCD}$ parameter which sets a scale of energy in the theory of
strong interactions (the only dimensional quantity in QCD); see e.g.
Weinberg (1996, Vol. II, p. 153-156). The phase transition temperature
is roughly $T = T_c \sim \Lambda_{QCD} \sim 0.2$ GeV $\sim 10^{12}$
K.\footnote{The precise prefactor has to be evaluated by numerical
calculations (it is not far from $\sim 1$) and it differs slightly for
the two phase transitions (chiral symmetry breaking and deconfinement)
which are investigated, e.g. in lattice simulations, in connection with
the formation of the quark-gluon plasma (see Laermann and Philipsen
(2003)).}

\ \\ {\em The Compton and the de Broglie wavelength as 
possible physical length scales above the quark-hadron transition} 

\noindent Bound systems, like atoms, set a (coarse) time scale through
transitions between energy levels, and a (finer, but more theory
dependent) length scale through their spatial extension. But are bound
systems really needed in order to have a rod (a length scale) which
could function e.g. in an Einstein light-clock (mentioned in the
beginning of section 4)? As other possibly relevant length scales,
physicists may point to the Compton and the de Broglie wavelength of the
elementary particles. Although these are physically motivated length
scales which are independent of the existence of bound systems, we
question whether these can function in the core of a clock (providing
physical processes with well-defined duration) in the very early
universe. 

Consider first the Compton wavelength. In and above the quark-gluon
phase, the quarks and leptons still possess the physical property of
mass. Thus, one may still have length scales if the Compton wavelength
$\lambda = \lambda_C = \hbar/(m c)$ of these particles can be taken to
set such a scale.\footnote{Note, that the Compton wavelength is not
related to the physical extension of the particle in question. For
instance, the electron (which has no known spatial extension) has a
larger Compton wavelength (since it has a smaller mass) than the proton
which has spatial extension of about 1 fermi $=10^{-15}$ meter. The
Compton wavelength rather sets a distance scale at which quantum field
theory becomes important since at those wavelengths there is enough
energy to create a new particle, $\hbar \omega_C = \hbar c/\lambda_C = m
c^2$. Thus, the Compton wavelength is the length scale at which `pair
production' of particle-antiparticle pairs occurs. (We should also note
that a force, e.g. the weak force, mediated by a massive particle is
effectively described by the Yukawa potential $V(r) \sim 1/r
\exp(-r/\lambda_C)$ which has a characteristic range set by the Compton
wavelength of the massive particle). } As discussed in the beginning of
section 4, however, the quantum field theoretical phenomenon of pair
production and the difficulties of precisely locating QED processes in
space and time make it difficult to maintain that the Compton wavelength
divided by $c$ corresponds to a physical process which can function as
the core of a clock.\footnote{Even if we grant that the Compton
wavelength is a length-scale setting physical process (that is, the
length scale at which pair production occurs) we note that since
$\lambda = \lambda_C = \hbar/(m c)$ is a scalar it will -- considered as
(some sort of) a core of a `measuring stick´ -- not be able to point in
any direction in 3-space (even in principle). Thus it will hardly provide a 
physical foundation for the concept of `pointing
in a direction' which appears to be necessary to give a physical
underpinning of the concept of `isotropy' (an input assumption in
building up the FLRW model in cosmology), see section 3.2.}

Apart from the Compton wavelength, a particle in quantum theory is
associated with a de Broglie wavelength $\lambda_{dB} = h/p$, where $h$
is Planck's constant and $p$ is the momentum of the particle (or wave).
Indeed, the de Broglie wavelength is a physical length scale for example
in a double slit experiment where the particle (e.g. an electron)
interacts with itself ($\lambda_{dB}$ can be inferred from the
interference pattern). In our assessment, however, one can hardly use
the de Broglie wavelength as the fundamental physical length scale
(which could function e.g. in an Einstein clock) in the early universe.
A first problem is that the definition of a de Broglie wavelength
depends on there being a fixed reference frame (with respect to which
the momentum can be specified), and part of our reason for asking for
length scales (rods) is precisely that they are needed as a physical
basis to set up the (comoving) coordinate frame. Setting up this
coordinate frame (in order to avoid a vicious circularity) should
therefore rest on other length scale setting processes besides the de
Broglie wavelength. Moreover, in the (comoving) coordinate frame the
material constituents (the particles) are at rest (in average). But in
the rest frame the momentum of the particle is zero, $p = 0$, and the de
Broglie wave length of the particle is thus infinite (undefined). This
makes the de Broglie wavelength unsuitable as a fundamental physical
length scale.\footnote{The length of any rigid rod is frame dependent in
special relativity. But a proper `measuring rod' should have a
well-defined length in the rest frame of the measuring rod (a
well-defined `rest length') and the rest de Broglie wavelength is
infinite.}
  
Moreover, in the ultrarelativistic limit, the rest mass of the particle
is negligible in comparison with the kinetic energy (so $E \approx
E_{\mbox{\tiny kin}} \approx pc$), and the particle behaves effectively
as a photon (that is, it is effectively massless).\footnote{The de
Broglie wavelength is proportional to $1/p$. Since the momentum grows
with temperature, it is a length scale which shrinks with temperature
(note that this is not the case for the length scales set by bound
systems and the Compton wavelength). Thus $\lambda_{dB}$ does not give a
fixed length scale relative to which an expanding cosmic scale factor, say,
could be gauged.} Thus, in the ultrarelativistic limit, the de Broglie
wavelength of a particle is analogous to the wavelength of a photon. As
we discussed in section 4.3, such a wavelength (or the corresponding
frequency) cannot by itself set a length scale, as electromagnetism is a
scale invariant theory.

\subsection {The electroweak phase transition: $\sim 10^{-11}$ seconds}

The standard model of the electroweak and strong interactions is a gauge
theory based on the general framework of quantum field theory (QFT). The
model embodies, at present, a Higgs sector (comprising a (set of)
postulated scalar field(s) $\phi$) which makes it possible to give
masses to various constituents of the standard model (invoking the idea
of spontaneous symmetry breaking) without destroying renormalizability
of the model. At low temperatures the quantum mechanical vacuum
expectation value of the Higgs field is non-zero.\footnote{This is one
of several sources in QFT of the so-called cosmological constant
problem, see e.g. Rugh and Zinkernagel (2002).} The electroweak phase
transition occurs, it is believed, at a transition temperature of $T
\sim 300$ GeV $\sim 10^{15}$ K when the universe was $\sim 10^{-11}$
seconds old. (Kolb and Turner 1990, p. 195). Above the phase transition
the Higgs field expectation value vanishes,
\begin{equation} \label{phizero}
 < \phi > \; = \; 0.
\end{equation}
This transition translates into {\em zero rest masses} of all the
fundamental quarks and leptons (and massive force mediators) in the
standard model. 

Thus, even if Compton wavelengths (corresponding to the non-zero rest
masses) could set a length scale below the electroweak phase transition
point, above the transition there are no Compton wavelengths for the
massless quarks, massless leptons and massless $W$'s and $Z^{0}$. Also,
the possibility to correlate the $t$-parameter with decay processes such
as the decay of e.g. the $Z^{0}$ particle is not possible above the
phase transition point since massless particles will not decay; see e.g.
Fiore and Modanese (1996).

In spite of the zero mass of the quarks and the leptons, there might
still be a rudiment of mass left in the model. Most of the mass
of a composite bound system like the pion (and also protons, which have
melted away at this temperature) does not arise from the $u$ and $d$
rest masses, but from the highly complicated gluonic self-interaction of
the QCD colour fields. Most of the pion mass, taken to be $\sim 0.6$ GeV
above the quark-hadron transition (D. Diakonov, personal communication),
might thus survive the transition where the fundamental quarks become
massless. However:

(1) It is questionable whether a pion survives as a bound system at
temperatures, $ T \gtrsim 300$ GeV, which are roughly $\sim$ 500 times
larger than its rest mass above the quark-hadron transition (and more
than one thousand times the QCD scale $\Lambda_{QCD} \sim 0.2$ GeV). 

(2) Even if a QCD bound system, like a pion, could survive as a bound
system at these high energies, a pion mass of $\sim 1$ GeV is to be
considered `effectively massless' relative to temperatures $T \gtrsim 300$
GeV. If a particle of rest mass $\sim 1$ GeV has an energy (kinetic
energy) of $\sim 300$ GeV it will move with a Lorentz $\gamma$-factor of
$\gamma \approx 300$ (ultrarelativistic limit). Thus the situation
resembles the problem -- discussed in section 4.3 -- of providing a
physical foundation for the concept of temperature $T \sim 300$ GeV from
a determination, in principle, of the frequency (or wavelength) of the
photon. The wavelength of a photon sets a length scale only when
reference can be made to microstructure (outside the theory of
scale-invariant electromagnetism). More quantitatively, a photon
wavelength of order $\sim \hbar c/ (k T) \sim 10^{-18}$ meter
(corresponding to temperatures $\sim 300$ GeV) should have its physical
basis in length scales $\sim 10^{-18}$ meter defined by microstructure
which is at rest relative to the comoving coordinate system.\footnote{In
our assessment the length scales should be length scales (`measuring
sticks') for physical structure which is at rest in the comoving cosmic
coordinate system. (Recall that the cosmic standard clocks and measuring
rods are at rest in the comoving coordinate system). Indeed, it seems
unreasonable to ground the concept of a physical scale on the fact that
a `measuring stick' (at a given length scale) which moves with velocities
arbitrarily close to $c$ can get arbitrarily Lorentz contracted. If one
were to allow Lorentz shortened objects as a basis for physical scales,
then the meter stick in Paris would be a physical basis for the Planck
scale if we consider it from a moving system with a $\gamma$-factor of
$10^{43}$!} The small non-zero rest mass $\sim 1$ GeV of a pion {\em
does} break the scale invariance which is a symmetry property of a
perfectly massless photon. But in our assessment the relevant length
scale of this symmetry breaking (the Compton wave length of a $\sim 1$
GeV pion is $\sim 0.2$ fermi $= 0.2 \times 10^{-15}$ meter) is
insufficient to provide a physical basis for the energies and
temperatures operating at scales of $\sim 300$ GeV. The pion `measuring
stick' is more than one hundred times too `coarse'!

In closing this brief discussion of length scales which may or may not
be established by the existence of a $\sim 1$ GeV rest mass particle, it
is instructive to compare the time and energy scales which are
associated with such a particle (of rest mass 1 GeV) and the time and
temperature (energy) scales of the universe at the times of the
electroweak phase transition. For the purpose of this comparison, we
assume that the Compton wavelength of a massive particle is able to set
a physical length scale (see also discussion in section 5.2): A rest
mass of 1 GeV then corresponds to length scales (via the Compton wave
length) of $\sim 0.2$ fermi $= 0.2 \times 10^{-15}$ meter, and time
scales of $\sim \hbar/(m c^2) \sim 10^{-24}$ seconds. Thus, this time
scale has a very fine resolution (is sufficiently fine-grained) relative
to the `age of the universe' ($\sim 10^{-11}$ seconds at the point of
the electroweak phase transition). On the other hand the same rest mass
of 1 GeV corresponds to energy and temperature scales of $\sim 1$ GeV
$\sim 10^{13}$ K which (given the argumentation above) appear
insufficient (by a factor of 300) to provide a physical basis of the
physics (e.g. the temperature concept) operating at energy and
temperature scales of $\sim 300$ GeV $\sim 3 \times 10^{15}$ K.

In the cosmological context, both time and temperature scales need to be
based in sufficiently fine-grained physical structure in order to yield
a physical basis for the time-temperature relation (such as equation
(\ref{timetemp})). (Recall that it is {\em temperature} which determines
the time of the phase transition, and thus -- not least -- temperature
which needs a physical basis).

\subsection{Nearly scale invariance above the electroweak phase
transition}

We now mention a significant problem which arises when we enter a phase
where the masses of the fundamental particles disappear. The question is
whether we can point to {\em any} relevant physical phenomena which set
a physical length scale (fine-grained or not) above the electroweak
phase transition where the universe is envisaged to contain
structureless constituents of massless fermions (leptons and quarks) and
bosons (electroweak photons and strongly interacting
gluons).\footnote{Note that a universe of only massless constituents
will be described in terms of pure radiation, and we are thus confronted
with a situation similar to the one in section 4.3 concerning the
blackbody radiation `clock', but in this case without any available
material length scales to physically interpret the frequency of the
radiation.}

The gauge theories of the standard model exhibit a symmetry (modulo some
small effects that we shall address in a moment) which implies a scale
invariance of the theories when the masses of the particles are
zero.\footnote{The symmetry (exhibited by a massless gauge theory) is
known as conformal invariance. Conformal symmetry is a stronger symmetry
than scale invariance: Scale invariance requires invariance under
uniform length rescaling, whereas conformal invariance also permits a
non-uniform local rescaling and only requires that angles are kept
unchanged. (The difference between scale and conformal symmetry is
roughly analogous to the difference between global and local gauge
symmetry).} The scale symmetry operation, which is in conformity with
special relativity (keeping $c$ invariant) and quantum
mechanics (keeping $\bigtriangleup \hat{x} \bigtriangleup \hat{p} \geq
\hbar$ and $\bigtriangleup t \bigtriangleup \hat{E} \geq \hbar$
invariant), is the set of transformations (with scaling parameter
$\lambda$),
\begin{equation} \label{conformalsymmetry}
t \rightarrow \lambda \; t \; \; , \; \; \vec{\BFACE{x}} \rightarrow
\lambda \; \vec{\BFACE{x}} \; \; , \; \; E \rightarrow 1/\lambda \; E \;
\; , \; \; \vec{\BFACE{p}} \rightarrow 1/ \lambda \; \vec{\BFACE{p}} 
\end{equation}
The transformation of other physical quantities under this scaling
operation follows from these transformation rules.\footnote{For example,
the concept of temperature transforms as energy, $T \rightarrow
1/\lambda \; T$, and the concept of energy density $\rho = E/ V$
transforms as $\rho \rightarrow (1/ \lambda^{4}) \; \rho$, etc. } 

The constants of the governing theory (the quantum field theory of gauge
interactions) such as $\hbar$ and $c$ are held invariant under the
transformations (\ref{conformalsymmetry}). Constants like the
electromagnetic fine structure constant $\alpha_e = e^2/ (\hbar c)$ are
also invariant under the transformations. If masses were present they
would have to be held constant (otherwise we would change the theory in
question), but the equations of the theory would then not be invariant
under the scale symmetry operations (\ref{conformalsymmetry}). In this
way the presence of masses in the theory break the
conformal invariance of the theory. However, above the electroweak phase
transition the Higgs expectation value vanishes (cf. eq.
(\ref{phizero})), which translates into zero rest masses of all the
fundamental quarks, leptons and massive force mediators in the standard
model.\footnote{In finite temperature QFT there are temperature
corrections to the rest masses, but the masses which `effectively' arise
in this way will be proportional to the temperature $T$ and will thus
not be able to set a fixed scale which can provide a physical basis for
$T$ itself. The Higgs mass itself has a slightly more complicated story.
Above the phase transition $m_{Higgs}^2$ has contributions from a
(tachyonic) mass term $m_0^2 < 0$ as well as the finite temperature $T$
contribution. It is unclear to us whether such a mass term can be
associated with some physical process which could function as the core
of a clock. In any case, well above the phase transition the effective
Higgs mass will (again) be proportional to the temperature $T$. So, in
the `desert' well above the phase transition the high-$T$
effective Higgs mass will not be able to provide a physical foundation
for the temperature concept itself.} The equations of the standard model
will then remain invariant under the conformal rescaling.

\begin{quotation}
\noindent
{\bf Observation $\# 4$:} {\em There is no physical basis for the time scale
if only conformally invariant material is available.} If
there is scale invariance of the available physics above the
electroweak phase transition it will be impossible to find physical
processes (among the microphysical constituents) which can set a fixed
scale. \end{quotation}

\noindent This would imply that we cannot even in principle find a core
of a clock among the microphysical constituents of the universe at this
early stage: The relevant physics (the electroweak and strong sector)
cannot set physical scales for time, scales for length and no scale for
energy. If there is no scale for length and energy then there is no
scale for temperature $T$. Metaphorically speaking, we may say that not
only the property of mass of the particle constituents `melts away'
above the electroweak phase transition but also the concept of
temperature itself `melts' (i.e. $T$ loses its physical foundation above
this transition point).

We remark that Roger Penrose (2006) has recently pursued ideas about
(the lack of) time in a conformally invariant universe which are similar
in spirit to our discussion (above and in previous versions of this
manuscript). Cf e.g. (2006, p. 2761):

\begin{quotation}
\noindent
With [...] conformal invariance holding in the very early universe, the
universe has no way of ``building a clock". So it loses track of the
scaling which determines the full space-time metric, while retaining its
conformal geometry.\newline [...] the universe ``forgets" time in the sense
that there is no way to build a clock with just conformally invariant
material. This is related to the fact that massless particles, in
relativity theory, do not experience any passage of time.
\end{quotation}                                
         
\noindent
In spite of the similarity with our discussion, however, Penrose appears
to assume the necessary link between time and clocks without providing
further argumentation.\footnote{Whereas we attempt in the present
manuscript to initiate a systematic study of the physical basis of time,
the context of the quote is Penrose's proposal of an ``outrageous new
perspective" (a conformal cyclic cosmology) in which approximate
conformal invariance holds in both ends (the beginning and the remote
future) of the universe.} 

In the following we mention some possible avenues which
should be pursued in order to introduce the concept of scale to this
seemingly scale invariant physics and thus possibly provide some
(micro-)physical underpinning of cosmic time beyond $\sim 10^{-11}$
seconds. We are acquainted with three mechanisms which can break the
perfect scale symmetry:
\begin{quotation}
\noindent
(1) The dimensionless coupling constants $\alpha = \alpha (\mu)$
run with the energy scale $\mu$ (quantum anomalies). \\
(2) Gravitational interactions break the scale invariance. \\
(3) New speculative physics may introduce length and time scales.
\end{quotation}

\noindent A detailed discussion of these options involves complicated
physics beyond the scope of the present manuscript.\footnote{We are
currently investigating such attempts to base the conception of time
beyond the electroweak scale in collaboration with H.B. Nielsen.} Thus
we only present some very brief remarks here.

As concerns (1), the running of the coupling constants (of the various
gauge theories) is an effect of higher-order corrections in quantum
field theory (a quantum `anomaly').\footnote{This running is extremely
small for the electromagnetic running coupling constant: $\alpha_e =
e^2/(\hbar c)$ runs from $1/137$ at zero energy density, to $\sim 1/128$
at the electroweak scale to $\sim 1/110$ at the Planck scale. The
running is more substantial for the strong coupling constant which is
large at low energy density (non-perturbative QCD) and small at high
energy densities (the regime of perturbative QCD), cf. the notion of
`asymptotic freedom'. However, the parameter $\Lambda_{QCD} \sim 0.2$
GeV which separates these regimes is more than a factor one thousand
smaller than the energies and temperatures in question (above the
electroweak phase transition point).} One might imagine scattering
experiments (thought experiments) in the early universe which could
determine various values of $\alpha$, and then one could imagine that
these values, together with theoretical knowledge about the running of
$\alpha = \alpha_g (\mu)$, could be turned around to extract the energy
scale $\mu$. However, our preliminary finding is that physical scales
will be needed as `input' (rather than being derivable as output) in
order to specify the set-up of such experiments. This suggests that the
running of coupling constants in itself is insufficient to provide a
physical foundation of scales in the very early universe. 

As concerns (2), the gravitational interaction stands out (as compared
to the three other coupling strengths for the electroweak and strong
interactions) since it has dimension: $G = 6.67 \times 10^{-11} m^3
kg^{-1} sec^{-2} $.  
According to general relativity, gravity couples to the total energy,
not just to rest mass, and it is standard to provide a dimensionless
measure of the strength of the gravitational coupling as given by 
\begin{equation} \label{runninggravcoupling}
\alpha_g = \alpha_g (\mu) = \frac{G \mu^2}{\hbar c^5}
\end{equation}
In this way a running of the (dimensionless) gravitational coupling
constant with the energy scale $\mu$ is introduced.\footnote{In contrast
to the gauge theory coupling constants, the running of the gravitational
coupling constant is a trivial one: The gravitational constant $G$ has
dimension, and thus runs with a change of scale (not as a result of
renormalization group equations within the framework of
QFT).} The coupling constant runs from the extremely small values
$\alpha_g (\mu = m_p c^2) \; \sim \; 10^{-40} $ at the energy scale of
the proton mass up to order unity at the Planck scale, $\mu = E_P =
(\hbar c^5 /G)^{1/2} \sim 10^{19}$ GeV. Since the strength of the
gravitational coupling constant is so extremely tiny at scales of $\sim
1$ GeV ($\alpha_g (\mu = m_p c^2) \sim 10^{-40}$), it is standard to
assert that the gravitational interaction is simply totally negligible
in the realm of elementary particle physics. 

Introducing the gravitational constant in the problem will of course
yield a scale (since $G$ carries a mass scale with it -- when it is
combined with $\hbar$ and $c$). However we are suspicious of the scales
(Planck scales) introduced by the way of constants of nature, if these
cannot be associated with possible or actual physical processes at the
time. Theories of quantum gravity are still highly speculative, and
moreover there are well-known problems of even identifying a suitable
time parameter in such theories. In any case it is expected that quantum
gravity effects are negligible at energy scales around the electroweak
phase transition point (and negligible well into the `desert' above this
phase transition). On the other hand, the {\em classical} gravitational
field couples to radiation in the sense that the expansion of the
universe (as described by a scale factor $R = R(t)$) yields
corresponding redshifts in the radiation field. However, if the universe
merely consists of radiation (and massless particles) with wavelengths
which expand at exactly the same rate as the overall scale factor, then
`expansion' appears to be a physically empty concept (without a physical
basis), cf. the discussion in section 4.1. Thus, if the microscopic
structure of matter is scale invariant it might not be possible to
obtain a length scale from the interplay of (scale invariant)
microphysics with gravity; cf. also the Eddington remark quoted in
section 2.

After this short discussion of the role of gravity in setting a length
scale, we turn in the following subsection to the third option for
breaking the scale invariance.

\subsection{Beyond the standard model}

The standard model is expected to be just some `effective' theory
arising from a more fundamental theory with built-in constants and
parameters referring to higher energies than the parameters of the
electroweak standard model (governing the electroweak phase transition
described in sec. 5.4). \\

\noindent {\em Massive X particles populating the `desert'? } \\ The
standard model involves roughly 20 unexplained `input' parameters. Grand
unified theories (with a unification mass scale $M_X$) have been
constructed with the purpose of explaining the values of some of these
input parameters. In the simplest possible SU(5) GUT theory the
experimental bound on the proton lifetime implies that the new $X$
particles in the GUT theory have masses $M_X \gtrsim 10^{14} - 10^{15}$
GeV (only 4 orders of magnitude below the Planck scale) and there will
be a `great desert' with no new physics between the scales of the
electroweak scales $\sim 300$ GeV and the $M_X$ scales (Froggatt and
Nielsen 1990, p. 47). Such a scenario with no new particles over 12
orders of magnitude in the roughly scale invariant physics is not good
news for finding physical processes comprising the cores of
clocks.\footnote{However, it has recently been (optimistically)
contemplated that the onset of string theory, and therefore of quantum
gravity, could occur already at a scale much lower (around $\sim 10^3
\mbox{GeV} \sim 1 $TeV) than the Planck scale. This would imply that the
time (and space) concept would be severely modified (if at all
meaningful) already at times corresponding to TeV scales (see e.g.
Antoniadis et al. (1998)).} Rather it implies a `great desert' (from
$\sim 300$ GeV to $\sim 10^{14}$ GeV) for which the remarks about scale
invariance (and the lack of physical length scales) of subsection 5.4
apply.\footnote{Other more complex grand unified theories can be
constructed (see, e.g., Collins et al. 1989, p. 164). Supersymmetric
model building is another proposal for physics in the desert in which
each particle is paired with a superpartner (yet to be observed), and
such models thus introduce plenty of new (speculative) particles.}

If the time scale concept is to be founded upon the introduction of
speculative massive particles, we have an example of the following:

\begin{quotation}
\noindent
{\bf Observation $\# 5$:}
The time scale above the electroweak phase
transition is purely speculative in the sense that it cannot be founded
upon an extrapolation of well known physics (due to scale invariance,
section 5.4) above the phase transition point. Thus time (the time
scale) will have to be founded on the introduction of new physics
(beyond the standard model of particle physics), and is in this sense as
speculative as the new (speculative) physics on which it is based.
\end{quotation}

\noindent
{\em Time of inflation: $\sim 10^{-34}$ seconds?} \\ Whereas inflation
in its original conception was tied to (speculative) GUT theories,
spontaneous symmetry breaking (phase transitions), etc. -- thus
providing a bridge between speculative high energy physics and cosmology
-- this link between inflation and particle theories has become weaker
in subsequent developments (see also e.g. Zinkernagel (2002)).
Inflationary models often introduce a species of massive scalar fields
$\phi$ and potentials $V(\phi)$ accompanied by various adjustable
parameters which are only weakly linked to known (or speculated)
physical processes. If the parameters of some inflationary model are
adjusted so the duration of the inflationary phase is `$10^{-32}$
seconds', say, it appears that such an interval of time is put into the
model -- by hand -- via the fine-tuning of various mathematical
parameters of the toy-model, and not set by time scales of microphysics.
It is our suspicion that such procedures provide a rather weak physical
basis for the time scale concept.\footnote{We note that it is sometimes
asserted that inflation starts $10^{-34}$ seconds `after' the big bang,
and, moreover, that inflation is contemplated to have observational
consequences. For instance, it is speculated that the nearly scale
invariant spectrum of cosmological perturbations observed in the cosmic
microwave background radiation (cf. the COBE and WMAP experiments) could
have been created during inflation as far back as the quoted $t \sim
10^{-34}$ seconds, and that those cosmological perturbations
subsequently propagated, almost uninterruptedly, through the hot early
universe plasma -- over an interval of time which spans 47 orders of
magnitude! -- up to the times of the release of the CMB $\sim 10^{13}$
seconds after the big bang.}

The space-time metric (usually) envisaged in the inflationary phase is
the de Sitter metric, which is a spatially homogeneous and isotropic
solution to the Einstein equations for an empty space with a vacuum
density (cosmological constant) $\Lambda$. Some researchers speculate
that the particle content of the universe was first created with the
inflationary reheating.\footnote{Cf. e.g. Linde (2004, p. 454): ``All
matter surrounding us was produced due to the decay of the scalar field
after inflation. [...] All matter in the universe was produced due to
{\em quantum} processes after the end of inflation".} In this case, one
will have to physically ground the time (and length) scale concept
`during' inflation on a scalar field (acting as a
cosmological constant). Whereas such ideas may introduce length scales
in the envisaged early inflationary universe, the ideas remain of course
speculative and the time scale concept based upon various physical
processes which may take place in the inflation `epoch' (e.g. the time
scales associated with various oscillations in the $\phi$ field) is
likewise speculative. (The introduction of a massive scalar (inflaton)
field $\phi$ bears some resemblance to the introduction of speculative X
particles mentioned above).

In the (much studied) class of inflationary models known as ``chaotic
inflation" (Linde (2004)) it is of interest to note that it is necessary
to require that the initial value of the inflaton field {\em exceeds}
the Planck scale (see e.g. Peacock (1999), p. 336-337)
$$ \phi_{start} \; \; \gg \; \; m_{P} = (\hbar c/G)^{1/2} \sim 10^{19}
\; \mbox{GeV}/c^2 $$
Thus, not only do such envisaged mathematical scenarios involve
extrapolations of our physics framework (quantum field theory and
general relativity) down to Planck scales -- but even substantially
beyond Planck scales and right through the `era' of quantum gravity --
where it is a widely held view that the concepts of space and time lose
their meaning.

\section{Summary and concluding remarks}

In section 1 and 2 we point out that cosmology implies a tremendous
extrapolation of our ordinary concept(s) of time (and space).
Nevertheless, we argue that the concept of time in cosmology, as well as
in ordinary language, should be understood in relation to physical
processes which can serve as (cores of) clocks. Our position is in
conformity with a {\em relationist} view of space-time in the tradition
of philosophers like Leibniz, although the relationism we defend is not
reductionist -- that is, we maintain that time and physical processes
which can serve as (cores of) clocks are equally fundamental. In the
cosmological context we attempt to motivate, within the scope of the
time-clock relation, the criterion that in order to interpret $t$ of the
FLRW model as time, that is, in order for cosmic time to have a physical
basis, it must be possible to construct a core of a clock out of the
physics envisaged in the various epochs of cosmic history. In
particular, we maintain that in order for the cosmic time {\em scale}
(used to indicate the onset and/or duration of the various epochs) to
have a physical basis, the physical process acting as a core of a clock
must have a well-defined duration and must be sufficiently fine-grained
to `time' the epoch in question.

In section (2 and) 3 we argue in favor of the necessity of rods and
clocks in the theories of relativity and cosmology (e.g. for setting up
the FLRW reference frame). We note in particular that the classical
gravitational field in general relativity, by itself, is not able to set
any length or time scales (the combination of $c$ and $G$ is
insufficient to set such scales). General relativity has to `go outside
its own borders' and obtain such scales from the material content
(providing the cores of rods and clocks). We indicate also how the
empirical adequacy, and the very formulation of, the simplifying
assumptions in standard cosmology (the Weyl principle and the
cosmological principle) depends on physically based scales.

In section 4 and 5 we analyze possible clock systems (physical processes
which may function as the cores of clocks) in the early universe and
find that they all depend on the existence of physically founded scales
(e.g. the spatial extension of bound systems). For instance, we see that
the concept of temperature -- which is the most commonly used time
indicator in the early universe -- rests on the existence of
sufficiently fine-grained and physically based length and energy scales.
We note that such physical scales -- in order to be non-speculative --
should be based on well-known physics available at the cosmological
epoch in question. This alone suggests that the cosmic time (scale)
concept becomes speculative `before' $10^{-11}$ seconds -- corresponding
to energies ($\sim 300$ GeV) indicating the current upper limit of known
physics (the standard model of particle physics). 

Speculative physics is needed not only because presently known physics
is too `coarse' -- and thus unable to provide sufficiently fine-grained
physical scales to ground concepts like time, length and temperature in
very `early' epochs. As we discuss in section 5, within known physics
the existence of {\em any} physical scale (fine-grained or not)
gradually becomes more questionable as the temperature increases in the
very early universe. In this regard, we emphasize two important
`moments': (i) above the `quark gluon' phase transition, at $\sim
10^{-5}$ seconds there are roughly no bound systems left (liberation of
quarks); and (ii) above the electroweak phase transition, at $\sim
10^{-11}$ seconds all the fundamental constituents of the standard model
become massless. When all masses are zero, the gauge theories describing
the material constituents and their interactions are (locally) scale
invariant, and thus the material content of the universe cannot set
physical scales for time, length and energy.\footnote{As mentioned in
section 5.4, some small quantum anomaly effects as well as the force of
gravity are known to break this scale invariance, and it will be
important to study if physical length scales (which refer to
contemplated physical processes in existence at the time) can be based
on those symmetry breaking effects (or if new speculative physics has to
be introduced in order to set appropriate physical length scales).} If
there are no scales for length and energy then there is no scale for
temperature $T$. We have formulated this observation in the following
metaphorical form: Not only the property of mass of the particle
constituents `melts away' above the electroweak phase transition but
also the concept of temperature itself `melts' (i.e. $T$ loses its
physical foundation above the phase transition point).

These results suggest that the necessary physical requirements for
setting up a comoving coordinate system (the reference frame) for the
FLRW model, and for making the $t \leftrightarrow \mbox{time}$
interpretation, are no longer satisfied above the electroweak phase
transition -- unless speculative new physics is invoked. The physical
requirements include the (at least possible) existence of (cores of)
rods and clocks as well as the validity of the Weyl postulate. If there
are no length scales set by the material constituents above the
electroweak phase transition then the concept of a rod (like that of a
clock) will become unclear. As concerns the Weyl postulate, this
involves the idea of non-intersecting particle trajectories, often
envisaged to be the world lines of galaxies which are at rest in the
comoving frame. For massless particles with random movement at velocity
$c$, the Weyl postulate is not satisfied for a typical particle,
although one may attempt to introduce fictitious averaging volumes
(which thus require a length scale) in order to create
as-close-as-possible substitutes for `galaxies which are at rest'.

The role of time is a main topic in contemporary studies of quantum
gravity and quantum cosmology. We nevertheless hope to have made
plausible in the foregoing that there are interesting problems with
establishing a physical basis for the concept of time, and in particular
for the concept of a time scale, in cosmology much (thirty orders of
magnitude) before theories of quantum gravity may come
into play.

If it is correct that time is necessarily related to actual or possible
physical processes in the universe, then what are the consequences for
early universe cosmology? Although our considerations suggest scepticism
concerning the reliability of the FLRW model before the electroweak
phase transition, we do not imply that early universe cosmology should
abstain from investigating mathematical models which could be relevant
for our understanding of cosmic history. Rather, our analysis suggests
that any speculative model for the very early universe imagined to be
operative at some specific `time' ought to include considerations
concerning what kind of physical processes could be taken to physically
ground the time (scale) concept. But given that these processes are
based on speculative physics, it should be admitted that the time
(scale) concept will be speculative as well. As a more general
conclusion, we hope that this study of time contributes to a
clarification of the physical foundations of modern cosmology.

\section*{Acknowledgements}

The first public presentation of the main ideas in this manuscript took
place at the European Science Foundation conference on philosophical and
foundational issues in spacetime theories at the University of Oxford in
2004. Since then, unpublished versions of the manuscript have been
circulating and presented at various conferences -- leading to several
improvements from the comments we have received along the way. It is a
pleasure to thank Holger Bech Nielsen for collaboration and many lively
discussions on topics of the manuscript. We are also grateful to Benny
Lautrup, Carl Hoefer, and Tomislav Prokopec for discussions and many
suggestions concerning earlier drafts. We thank anonymous referees for
constructive criticisms of earlier versions of the present work, and we
gratefully acknowledge the hospitality provided by the Center for the
Philosophy of Nature and the Niels Bohr Archive under the auspices of
Claus Emmeche and Finn Aaserud. One of us (HZ) thanks the Spanish
Ministry of Education and Science (Plan Nacional project BFF2002-01552
and HUM2005-07187-C03-03) for financial support.

\section*{References}

Alexander, H.G. (1956). {\em The Leibniz-Clarke Correspondence.} New
York: Philosophical Library.

Antoniadis, I., Arkani-Hamed, N. and Dimopoulos, S. (1998). New dimensions
at a millimeter to a Fermi and superstrings at a TeV.
{\em Physics Letters, B 436}, 257-263. hep-ph/9803315.

Audoin, C. and Guinot, B. (2001). {\em The Measurement of Time - Time,
Frequency and the Atomic Clock}. Cambridge: Cambridge University Press.

Barbour, J. (1989). {\em Absolute or Relative Motion?: A Study from a Machian
Point of View of the Discovery and the Structure of Dynamical Theories}.
Cambridge: Cambridge University Press.

Barbour, J. (1999). {\em The End of Time}. New York: Oxford University Press.

Bohr, N. and Rosenfeld, L. (1933). Zur Frage des Messbarkeit der
Elektromagnetischen Fieldgr\"{o}ssen. {\em Mat.-fys. Medd. Dan. Vid. Selsk.,
12}, no. 8. Trans. in J.A. Wheeler and W.H. Zurek (Eds.), (1983). {\em
Quantum theory and measurement} (pp 465-522). Princeton, NJ: Princeton
University Press.

Brown, H. (2006). {\em Physical Relativity}. New York: Oxford University
Press.

Coles, P. and Lucchin F. (1995). {\em Cosmology - The Origin and
Evolution of Cosmic Structure.} West Sussex: Wiley.

Coles, P. (ed.) (2001). {\em The Routledge Companion to: The New
Cosmology}. London: Routledge.

Collins, P.D.B.,  Martin A.D., and Squires E.J. (1989). {\em
Particle Physics and Cosmology}. New York: Wiley \& Sons.

Cooperstock, F.I., Faroni V., and Vollick D.N. (1998). The influence of
the cosmological expansion on local systems. {\em Astrophysical Journal,
503}, 61-66. 

Earman, J. and Norton, J. (1987). What Price Spacetime Substantivalism?
The Hole Story. {\em The British Journal for the Philosophy of Science,
38}, 515-525.

Eddington, A.S. (1939). {\em The Philosophy of physical science}.
Cambridge: Cambridge University Press. 

Eidelman et al. (2004). The Review of Particle Physics, Summary Tables
of Particle Properties. {\em Physics Letters, B 592}, 1-1107.
(Available at http://pdg.lbl.gov).

Einstein, A. (1920). {\em Relativity: The special and general theory}.
(Chapter 28). London: Routledge Classics.

Einstein, A. (1949). Autobiographical notes. In P.A. Schilpp (ed.), {\em
Albert Einstein: Philosopher-Scientist}. La Salle Illinois: Open Court.

Ellis, G.F.R. (2006). Issues in the Philosophy of Cosmology. In J.
Butterfield and J. Earman (eds.) {\em Handbook in Philosophy of
Physics}. (pp. 1183-1286). Amsterdam: North Holland. ({\em
astro-ph/0602280})

Feynman, R.P., Leighton, R.B. and Sands, M. (1963). {\em The Feynman
Lectures on Physics, Vol. 1}. New York: Addison-Wesley.

Fiore, G. and Modanese, G. (1996). General properties of the decay amplitudes
for massless particles. {\em Nuclear Physics B477}, 623-651.
({\em hep-th/9508018v4})

Frigg, R. and Hartmann, S. (2006). Models in Science. {\em The Stanford
Encyclopedia of Philosophy} (Spring 2006 Edition), Edward N. Zalta
(ed.),\\
http://plato.stanford.edu/archives/spr2006/entries/models-science/

Froggatt,C.D. and Nielsen, H.B. (1991)  {\em Origin of symmetries}.
World Scientific.

Gale, R.M. (1968). Introduction. In R.M. Gale (ed.), {\em The Philosophy
of Time -- A collection of essays} (pp. 1-8). London: Macmillan.

Gr{\" u}nbaum, A. (1973). {\em Philosophical problems of space and time
-- second enlarged edition}. Dordrecht: Reidel.

Gr{\" u}nbaum, A. (1977). Absolute and relational theories of space and
space-time. In J.S. Earman, C.N. Glymour and J.J. Stachel (eds.), {\em
Foundations of Space-Time Theories} (pp. 303-373). Minneapolis:
University of Minnesota Press.

Hackman, C. and Sullivan, D.B. (1995). Resource Letter: TFM-1: Time and
frequency measurement. {\em American Journal of Physics, 63}, 306 - 317.

Hoefer, C. (1996). The metaphysics of space-time substantivalism. 
{\em The Journal of Philosophy, XCIII}, 5-27.

Kolb, E.W. and Turner, M.S. (1990). {\em The Early Universe}.
New York: Addison-Wesley.

Laermann, E. and Philipsen, O. (2003). The status of lattice QCD at
finite temperature. {\em Annual Review of Nuclear and Particle Science,
53}, 163-208. hep-ph/0303042.

Linde, A. (2004). Inflation, quantum cosmology, and the anthropic
principle. In J.D Barrow, P.C.W. Davies and C.L.Harper (eds.) {\em
Science and Ultimate Reality}. Cambridge: Cambridge University Press.

Lopez-Corredoira, M. (2003). Observational Cosmology: caveats and open
questions in the standard model. {\em Recent Research Developments in
Astronomy \& Astrophysics, 1}, 561- 587. astro-ph/0310214

Marzke, R.F. and Wheeler, J.A. (1964). In H.Y. Chiu, and W.F. Hoffman
(eds.) {\em Gravitation and Relativity}. New York: Benjamin.

McLerran, L. (2003), RHIC Physics: The Quark Gluon Plasma and The Color Glass
Condensate: 4 Lectures, hep-ph/0311028.

Misner, C.W. (1969). Absolute zero of time. {\em Physical Review 186},
1328-1333.

Misner, C.W., Thorne, K., and Wheeler, J.A. [MTW] (1973). {\em
Gravitation}. New York:  W. H. Freeman

M\"{u}ller, B. and Srivastava, D.K. (2004), How Relativistic Heavy Ion
Collisions Can Help Us Understand the Universe, nucl-th/0407010.

Narlikar, J.V. (2002). {\em An Introduction to Cosmology}, (third
edition). Cambridge: Cambridge University Press.

North, J.D. (1990). {\em The measure of the universe -- A history of
modern cosmology}. New York: Dover.

Newton-Smith, W. H. (1980). {\em The Structure of Time}. London:
Routledge \& Kegan Paul.

Peacock, J.A. (1999). {\em Cosmological Physics}. Cambridge: Cambridge
UNiversity Press.

Penrose, R. (2006). {\em Before the Big Bang:
An outrageous new perspective and its implications for particle physics}. 
Proceedings of EPAC 2006, Edinburgh, Scotland, 2759-2767.

Pietronero, L. and Labini F.S. (2004). Statistical physics for complex
cosmic structures. {\em astro-ph/0406202}.

Poincar\' e, H. (1905). The Measure of Time. In H. Poincar\' e {\em The
value of science} (1958). New York: Dover.

Robertson, H.P. (1933). Relativistic Cosmology. {\em Review of Modern
Physics, 5}, 62-90. 

Rovelli, C. (1995). Analysis of the Distinct Meanings of the Notion of
``Time" in Different Physical Theories. {\em Il Nuovo Cimento, 110B},
81-93.

Rovelli, C. (2001). Quantum spacetime: What do we know? In C. Callender
and C. Huggett (eds.), {\em Physics meets philosophy at the Planck
scale: Contemporary theories in quantum gravity}. (pp. 101-124).
Cambridge: Cambridge University Press.

Rugh, S.E. and Zinkernagel, H. (2002). The quantum vacuum and the
cosmological constant problem. {\em Studies in History and Philosophy of
Modern Physics, 33}, 663-705.

Rugh, S.E. and Zinkernagel, H. (2008). Time and the cosmic measurement
problem. In preparation

Ryckman, T. A. (2001). Early Philosophical Interpretations of General
Relativity. {\em The Stanford Encyclopedia of Philosophy (Winter 2001
Edition)} Edward N. Zalta (ed.), 
http://plato.stanford.edu/archives/win2001/entries/genrel-early/.

Schramm, D.N. and Turner, M.S. (1998), Big-bang nucleosynthesis enters
the precision era, {\em Rev.Mod.Phys.} {\bf 70}, 303-318.

Schweber, S.S. (1997). The Metaphysics of Science at the End of a Heroic
Age. In R.S. Cohen, M. Horne and J. Stachel (eds.) {\em Experimental
Metaphysics}. (pp. 171 - 198). Dordrecht: Kluwer Academic Publishers.

Smolin, L. (2003). Time, structure and evolution in cosmology. In A.
Ashtekar {\em et al} (eds.), {\em Revisiting the Foundations of
Relativistic Physics -- Festschrift in Honor of John Stachel} (pp.
221-274). Dordrecht: Kluwer.

Stueckelberg, E.C.G (1951). Relativistic quantum theory for finite time
intervals. {\em Physical Review 81,1}, 130-133.

Trodden, M. (1998). Electroweak baryogenesis: A brief review.
hep-ph/9805252.

Weinberg, S. (1972). {\em Gravitation and Cosmology.} New York: Wiley and
Sons.

Whitrow G.J. (1980). {\em The natural philosophy of time.} Oxford:
Clarendon Press.

Whitrow, G.J. (2003). {\em What Is Time?} Oxford: Oxford University
Press.

Zinkernagel, P. (1962). {\em Conditions for description}. London:
Routledge and Kegan Paul.

Zinkernagel, P. (2001). {\em The Power of Customary Views, Part 2,
Documentation.} Copenhagen: Gerd Preisler Publishing.

Zinkernagel, H. (2002). Cosmology, particles, and the unity of science.
{\em Studies in History and Philosophy of Modern Physics, 33}, 493-516.

Zinkernagel, H. (2008). Did time have a beginning?. Forthcoming in {\em
International Studies in the Philosophy of Science}.


\end{document}